\documentstyle[pre,aps,floats,epsfig]{revtex}

\begin{document}



\title {Coefficient of restitution for elastic disks
}
\author{Franz Gerl and
        Annette Zippelius}

\address{
Institut f\"ur Theoretische Physik,
Universit\"at G\"ottingen,      \\
Bunsenstrasse 9, D-37073 G\"ottingen, Germany}

\date{\today}

\maketitle

\topmargin-0.5cm

\begin{abstract}
  We compute the coefficient of restitution, $\epsilon$, starting from
  a microscopic model of elastic disks. Collisions are found to be in
  general inelastic with a finite fraction of translational energy
  being transfered to elastic vibrations. The coefficient of
  restitution depends on the relative velocity of the colliding disks,
  such that collisions become increasingly more inelastic for larger
  relative velocities. The predictions of the model are in agreement
  with the approach of Hertz in the quasistatic limit.  The
  coefficient of restitution depends on the elastic constants of the
  material via Poisson's number. The elastic vibrations absorb kinetic
  energy more effectively for soft materials with low ratios of shear
  to bulk modulus.
\end{abstract}

\section{Introduction}

The most important characteristic of granular media is
the inelastic nature of the interparticle collisions.
The removal of kinetic energy in a granular gas
is responsible for nonequilibrium phenomena that are
of theoretical and experimental interest.
In computer simulations of granular matter this energy
loss is usually treated in a very simplified way.
In event driven simulations~\cite{ED} a fixed
coefficient of restitution is used, i.e.~after each
interparticle collision a certain fraction of the energy
involved is lost.  
In experiment~\cite{Experiment} the coefficient of restitution is found
to depend on velocity.
When using molecular dynamics techniques~\cite{MD}
ad hoc phenomenological assumptions for the intergrain force laws 
are introduced. (For a recent review, see \cite{Wolf}.)

Several microscopic mechanisms for the decay of kinetic energy during
collisions have been discussed.
Permanent plastic deformation of granular particles has been proposed
as a possible mechanism for the removal of kinetic energy 
\cite{Andrews}.
Viscoelastic behaviour  was used to extend the theory of Hertz
\cite{Hertz} to inelastic impact. One either invokes
a phenomenological damping term in the equations of
motion \cite{Poeschel} or uses a quasistatic approximation
for low relative impact velocities \cite{Brilliantov}.

More recently temporary storage of energy in elastic 
modes \cite{Goetz,Timo} has been discussed for one-dimensional
rods. During collisions energy is exchanged between translational
motion and internal degrees of freedom, whereas in
between collisions the energy stored in the elastic modes decays. 
If the collision rate is high enough, which is frequently observed in
simulations as a precursor for inelastic collapse, 
elastic energy may be transformed back into kinetic energy.
Thereby inelastic collapse is avoided and 
rich dynamics of temporary clusters is observed, including 
breakup and reformation of clusters as well as relaxation to a final
state with all particles in contact~\cite{Timo2}.

The quantity controlling the fraction of kinetic energy, which is
lost in a collision, is the coefficient of restitution $\epsilon$.
We will restrict ourselves to head-on, normal collisions,
where $\epsilon$ is simply defined as the ratio of relative velocities after 
and before collision
\begin{equation}
\epsilon:= -  v_{\rm f}/v_{\rm i}  \;\: .
\end{equation}
For perfectly elastic {\it one-dimensional} rods phenomenological wave
theory yields $\epsilon=l_1/l_2$, independent of the relative velocity
of the colliding rods. Here $l_1$ $(l_2)$ denotes the length of the
shorter (longer) rod. If both rods are identical we have $\epsilon=1$
always.  Using the approach of Hertz, Rayleigh \cite{Rayleigh} has
estimated, that the fraction of energy stored in the fundamental mode
for two slowly colliding spheres is about $0.02 \cdot v_{\rm i}/c$ 
(where $c$ is the velocity of sound in a one-dimensional
rod of the same material).  While
being quite small for many realistic scenarios his value for the
coefficient of restitution thus depends on velocity and does not
vanish for identical spheres.

In this paper we shall analyze collisions of two-dimensional elastic
disks.  Starting from three-dimensional elastic objects, the
two-dimensional case can be realized in two ways, called plane stress
and plane strain~\cite{Timoshenko}.
Plane stress describes the situation of thin
plates, where the stress components on the faces of the plates
disappear.  With the additional assumption of in-plane oscillations
only, the equations of two-dimensional elasticity apply.  In the case
of plane strain the end sections of a prismatic body are confined
between smooth rigid planes, so that displacements in the axial
direction are prevented. If the forces do not vary along the length
for symmetry reasons, it may be assumed that there is no axial
displacement anywhere.  This again reduces the problem to two
dimensions.
It is easy to show that the equations for plane strain are the same as
those for plane stress, provided one makes the substitution
\begin{equation}
\label{scal}
E \to \frac E{1-\nu^2} \; , \qquad \nu \to \frac \nu{1-\nu} \;
\end{equation}
where $E$ denotes Young's modulus and $\nu$ Poisson's number.  For
plane stress Poisson's number $\nu$ has the three dimensional value
and hence is restricted to $-1\le \nu\le 1/2$. With the above
transformation this corresponds to $-1/2\le \nu \le 1$ for plane
strain. In particular values of $\nu$ close to 1 can be achieved 
for plane strain by
materials with low shear modulus $\mu$, like rubber. 

As in our previous work for one-dimensional elastic
rods~\cite{Goetz,Timo}, we model the
transfer of translational energy to elastic vibrations of
two-dimensional disks and discuss the collision between two identical
disks or equivalently the collision of one disk with a rigid wall.
The interaction is modelled by a hard core potential depending on the
instantaneous separation between the two disks. In contrast to the
one-dimensional case the equations of motion for the elastic vibrations
have to be solved numerically. The main result of our paper is the
coefficient of restitution as a function of initial
relative velocity and elastic properties of the disks, as
shown in Figure \ref{Zweivgl}. Collisions are found to be elastic
for vanishing relative velocity in agreement with Hertz' quasistatic
approach \cite{Hertz}
and become increasingly inelastic for increasing relative
velocity. For soft matter, characterized by a small value of the shear
modulus, the elastic modes provide a rather effective mechanism for
the uptake of kinetic energy during collision, so that the
coefficient of restitution decreases with decreasing shear modulus.
In contrast to one dimension, where the collision of two rods of equal
length is always perfectly elastic \cite{Timo}, we find that the
collision between two identical disks is in general inelastic.

 Our paper is organized as follows. In section \ref{Moden} we discuss
two-dimensional elasticity and calculate the eigenmodes of a disk with
force free boundaries. In section
\ref{Koeffs} we  analyze the equations of motion of an elastic
disk, colliding with a hard wall. The numerical solution of the
equations of motion are discussed in section \ref{Numerik}. In section
\ref{Static} we  apply Hertz' and Rayleigh's
methods to two dimensions to study the limiting case
of very small relative velocities, i.e. the quasistatic limit.
We conclude with a summary of our results and an outlook.

\section{Elastic extensional modes of disks}
\label{Moden}
\subsection{Elasticity of planar bodies}

We briefly review the theory of linear elasticity in two dimensions,
as discussed e.g. by Love \cite{Love}, 
Landau \cite{Landau} or Timoshenko and Goodier
\cite{Timoshenko}.
We consider small displacements ${\bf u}({\bf x})$ and expand the free
energy up to quadratic order in the strain field, which is defined as
the symmetrized derivative of the displacement vector:
\begin{equation}
\label{VerschVerz}
\epsilon_{ij}= \frac 12 \left(\frac{\partial u_i} {\partial x_j} + 
\frac{\partial u_j}{\partial x_i}
\right) \;\;.
\end{equation}
An isotropic elastic body has only two independent elastic constants,
the shear modulus $\mu$ and the compressional modulus $K$. In terms of
these the elastic free energy is given by
\begin{equation}
E_{\rm elas}=\int d^2x\left(\mu ( \epsilon_{ij}-\delta_{ij}\epsilon_{ll}/2)^2
+K\epsilon_{ll}^2/2 \right).
\end{equation}
The notation implies summation over indices which appear twice.
The stress tensor is defined as the derivative of the elastic free
energy with respect to the strain field and explicitly given by
\begin{equation}
\label{stress}
\sigma_{ij}=\frac{\delta E_{\rm elas}}{\delta \epsilon_{ij}}=
2\mu(\epsilon_{ij}-\delta_{ij}\epsilon_{ll}/2)+K\delta_{ij}\epsilon_{ll}
\; .
\end{equation}
It is sometimes convenient to introduce Young's modulus $E$ and 
Poisson's number $\nu$ instead of the moduli of torsion and
compression. These quantities are simply related, in two dimensions we
have $E=(4\mu K)/(K+\mu )$ and $ \nu=(K-\mu)/(K+\mu)$,
so that alternatively to Eq.~(\ref{stress}) we may use
\begin{equation}
\label{stress.strain}
\sigma_{ij}=\frac{E}{(1-\nu^2)}((1-\nu)\epsilon_{ij}+\nu 
\delta_{ij}\epsilon_{ll}).
\end{equation}
In the absence of body forces the equations of equilibrium read
\begin{equation}
\frac{\partial \sigma_{ij}}{\partial x_j} =0 \ .
\end{equation}

\subsection{Normal modes}

In the following we shall discuss the eigenmodes of a circular disc
with force free boundaries (see also
\cite{Love}). Our starting point are the linear
equations of motion \cite{Landau}

\begin{eqnarray}
  \label{Bewgleich} {\cal L}{\bf u}:=
\frac{1+\nu}{2}\nabla(\nabla\cdot{\bf u})+\frac{1-\nu}{2}\Delta{\bf u}
& = & \frac{\varrho(1-\nu^2)} {E} \, 
\frac{\partial^2 {\bf u}}{\partial t^2} \nonumber \\
\end{eqnarray}
where $\rho$ is the mass per unit area. The differential operator 
${\cal L}$ is hermitean for force free
boundary conditions, as shown in the Appendix.
To solve the coupled differential equations for $u_x$ and $u_y$, we
introduce the areal dilatation $\eta$ and rotation $\zeta$
\begin{equation}
\eta = \frac{\partial u_x}{\partial x} + \frac{\partial u_y}
{\partial y}  , \nonumber \\ 2 \zeta = \frac{\partial u_y}{\partial x}
-\frac{\partial u_x}{\partial y} \;\; .
\end{equation}
The equations of vibration (\ref{Bewgleich}) then simplify to
\begin{eqnarray}
  \label{Simpbew1}
  \frac{\partial \eta} {\partial x} -(1-\nu) \frac{\partial   \zeta} 
{\partial y} &=& \frac {\rho(1-\nu^2)}{E}
\frac{\partial^2 u_x}{\partial t^2} \; ,  \\  \label{Simpbew2}
\frac{\partial \eta}{\partial y} +(1-\nu) \frac{\partial   \zeta}
 {\partial x} &=& \frac {\rho(1-\nu^2)}{E}
\frac{\partial^2 u_y}{\partial t^2}    \; .
\end{eqnarray}
Differentiating (\ref{Simpbew1}) with respect to $x$, (\ref{Simpbew2})
with respect to $y$ and adding them yields
\begin{equation}
  \label{Vibgl1}
  \nabla^2\eta= \frac {\rho(1-{\nu}^2)}E \frac{\partial^2
    \eta}{\partial^2 t}\;.
\end{equation}
Differentiating (\ref{Simpbew1}) with respect to $y$, (\ref{Simpbew2})
with respect to $x$ and subtracting them gives
\begin{equation}
  \label{Vibgl2}
 \nabla^2\zeta=   \frac {2\rho(1+\nu)}E \frac{\partial^2 \zeta} 
{\partial^2 t} \; .
\end{equation}
Hence, the equations of two-dimensional elasticity have been reduced to
two scalar wave equations which are coupled through the boundary conditions.

To discuss the eigenmodes of a circular disk, we use polar coordinates 
$r,\phi$ with the origin at the center of the disk. The
two-dimensional displacement vector is written as 
${\bf u}=u_x{\bf e}_x+u_y{\bf e}_y=u_r{\bf e}_r+u_{\phi}{\bf
  e}_{\phi}$. The radial  and tangential  displacements $u_r$ and $u_{\phi}$
are given by
\begin{equation}
  \label{Poldisp}
  u_r= u_x \cos \phi + u_y \sin \phi\;, \qquad u_{\phi}=- u_x\sin \phi
  + u_y \cos \phi \; .
\end{equation}
Similarly strain, dilatation and rotation can be worked out in polar
coordinates 
\begin{eqnarray}
\label{polar}
\eta= \frac{\partial u_r}{\partial r} + \frac{u_r}{r} +\frac 1r
\frac{\partial u_{\phi}}{\partial \phi} \;, \qquad
2 \zeta = \frac{\partial u_{\phi}}{\partial r} + \frac{u_{\phi}}{r} -\frac 1r 
\frac{\partial u_r}{\partial \phi} \; ,\\
\epsilon_{rr}=\frac{\partial u_r}{\partial r} \;, \qquad
\epsilon_{\phi \phi}=\frac{u_r}{r} +\frac 1r
\frac{\partial u_{\phi}}{\partial \phi} \;, \qquad
2\epsilon_{r \phi}=\frac{\partial u_{\phi}}{\partial r} -
\frac{u_{\phi}}{r} +\frac 1r 
\frac{\partial u_r}{\partial \phi} \; . \nonumber
\end{eqnarray}
Finally, because the medium is isotropic the stress-strain relation
in polar coordinates is as simple as it is for cartesian coordinates 
Eq.~(\ref{stress.strain}):
\begin{eqnarray}
\label{Kompression}
\sigma_{rr}\:=\:\frac{E}{1-\nu^2} (\epsilon_{rr}+\nu \epsilon_{\phi
  \phi}) \;, \qquad \sigma_{\phi \phi} \:=\:\frac{E}{1-\nu^2} 
(\epsilon_{\phi \phi}+\nu \epsilon_{rr}) \;, \qquad
\sigma_{r\phi}= \frac {E} {2(1+\nu)} \epsilon_{r\phi} 
\label{Scherung} \; .
\end{eqnarray}

The solutions to (\ref{Vibgl1}) and (\ref{Vibgl2}) are of the form 
\begin{equation}
\label{symsol}
\eta=A_n \kappa^2 J_n(\kappa r)\cos n\phi \, \cos \omega t \;, \qquad 
2\zeta= B_n \kappa'^2 J_n(\kappa' r) \sin n\phi \, \cos \omega t
\; ,
\end{equation}
where $J_n$ is Bessel's function of order $n$ \cite{Abramowitz}, $A_n$
and $B_n$ are constants and
\begin{equation}
\label{kapDef}
\kappa = \omega\, \sqrt{\frac {(1-\nu^2)\rho} E}, \qquad 
\kappa'= \omega\,\sqrt{\frac{2(1+\nu)\rho}E} \; .
\end{equation}
There is a second set of solutions with ``$\cos$'' and ``$\sin$''
interchanged. We shall only consider situations which are symmetric 
in $\phi$, so that $u_r(r,\phi)=u_r(r,-\phi)$  and
$u_{\phi}(r,\phi)=-u_{\phi}(r,-\phi)$. Hence we restrict ourselves to
the above set. The displacements corresponding to the above solutions
are given by 
\begin{eqnarray}
\label{eigen.u}
u_r(r,\phi)&=&  \left[A_n \frac{{\rm d}J_n(\kappa r)}{{\rm d}r} + nB_n
  \frac{J_n (\kappa' r)}{r}\right]  \cos n \phi  \, \cos \omega t\; , \\
u_{\phi}(r,\phi)&=& -\left[nA_n  \frac{J_n(\kappa r)}{r} + B_n
            \frac{{\rm d}J_n(\kappa' r)}{{\rm d}r}  \right] 
            \sin n\phi \, \cos \omega t \; . \nonumber
\end{eqnarray}
This can be easily verified by substituting Eqs.~(\ref{eigen.u}) into 
Eqs.~(\ref{polar}),
thereby one recovers the symmetric solution of Eq.~(\ref{symsol}).

At the boundary of the disk, $r=R$, shear and compression have to
vanish: $\sigma_{rr}(R,\phi)=0$ and $\sigma_{r\phi}(R,\phi)=0$. 
For $n=0$ and $B=0$ we have purely radial vibrations in which $u_{\phi}$ 
vanishes and $u_r$ is independent of $\phi$. The boundary condition
then reads
\begin{eqnarray}
\label{Radialvib}
\frac{{\rm d} J_1(\kappa R)}{{\rm d}R} + \frac \nu R J_1(\kappa
  R) =0 \;\;.
\end{eqnarray}
Areal dilatation and stress are maximal at the center of the disk,
although the center is not displaced. The displacemnet vanishes along
node lines, which are circles for $n=0$. 

There is a second set of solutions with $u_r=0$, $u_\phi$ independent
of $\phi$ and node lines which are cirles.  These modes ar not excited
in head-on collsions for which we impose the symmetry
$u_\phi(r,\phi)=-u_\phi(r,-\phi)$.

In the general case, i.e. for $n\ge 1$, there is no tangential
displacement, $u_\phi = 0$, for $\phi=\frac {\pi k}n$ and no radial
displacement, $u_r=0$ for $\phi=\frac{\pi(2k+1)}{2n}$ with $ k\in
\{0\ldots 2n-1\}$. These values of $\phi$ represent $n$ node lines for
either $u_\phi$ or $u_r$. The boundary conditions in general 
imply two equations
\begin{eqnarray}
-A_n \left[\frac{1-\nu}R \frac{{\rm d} J_n(\kappa R)}{{\rm d}R}
+ \Big(\kappa^2-\frac{1-\nu}{R^2}n^2\Big) J_n(\kappa R)
 \right] \qquad \qquad \quad \nonumber\\ \label{Anfvollst}
 + \quad nB_n(1-\nu) \left[
\frac 1R \frac{{\rm d} J_n(\kappa' R)}{{\rm d}R}
- \frac 1{R^2} J_n(\kappa' R)
\right] \;=\;0  \quad ,\\\nonumber   \\ \nonumber
-2nA_n \left[\frac 1R \frac{{\rm d} J_n(\kappa R)} {{\rm d} R} -\frac 1{R^2} 
J_n(\kappa R) 
\right] \qquad  \qquad \qquad \qquad \qquad  \qquad\\ + \quad  B_n \left[
 \frac 2R \frac{{\rm d} J_n(\kappa' R)}  {{\rm d} R}+ 
\Big(\kappa'^2-\frac{2n^2}{R^2}\Big) J_n(\kappa' R)
\right] \;=\; 0 \quad .  \label{Anfvollst2}
\end{eqnarray}
We eliminate $A_n$ and $B_n$ and use 
\begin{equation}
\frac{{\rm d} J_n(\kappa R)} {{\rm d} R} = \kappa J_{n-1}(\kappa R) - n 
\frac {J_n(\kappa R)} R
\end{equation}
to obtain the eigenvalue equation
\begin{eqnarray} \nonumber
(1-\nu)^2 (1-n^2) \kappa \kappa' J_{n-1}(\kappa R)J_{n-1}(\kappa' R) \qquad\\
+(1-\nu)\left(\kappa^2-(1-\nu)(1-n^2)n\right)\left(\kappa
  J_{n-1}(\kappa R)J_n(\kappa' R)
+ \kappa' J_{n-1}(\kappa' R)J_n(\kappa R) \right) \nonumber \qquad \nonumber\\
+ \left(\kappa^4-2(n^2+n)(1-\nu)\kappa^2\right) J_n(\kappa R)
J_n(\kappa' R) =0
\; . \label{Zero}
\end{eqnarray}

For any fixed number $n$, there are infinitely many
solutions $\omega_{n,l}$, numbered by $l=0,\ldots ,\infty$. 
These solutions have been determined numerically. Sometimes 
solutions for neighbouring $l$ are very closely spaced and
difficult to find numerically. It is then advantageous to
simultaneously determine the zeros of the derivative of (\ref{Zero}),
which are much more regularly spaced. In between a pair of zeros of
the derivative, we then search for a zero of the function.

The ratio $A:B$  is determined by reinserting the values
for $\omega_{n,l}$ into (\ref{Anfvollst}). 
We are left with one free
constant for each eigenmode. This constant can be fixed by using
normalized eigenmodes, according to
\begin{eqnarray}
\label{Normdef}  
\int\limits_0^{2\pi} {\rm d}\phi \int\limits_0^R 
{\rm d}r \: r ((u_r^{n,l}(r)\cos(n\phi))^2+(u_{\phi}^{n,l}(r)\sin(n\phi))^2)
= \pi R^2\;.
\end{eqnarray}
Here we have introduced
dimensionless quantities $u_r^{n,l}(r)$ and
$u_{\phi}^{n,l}(r)$ for the radial variation
the displacement (Eq.~\ref{eigen.u}):
\begin{eqnarray}
\qquad R\, u_r^{n,l}(r)&:=& \left[A_{n,l}
  \frac{{\rm d} J_n(\kappa_{n,l}r)}  {{\rm d} r}+ nB_{n,l} 
\frac{J_n(\kappa'_{n,l}r)}r\right]\; , \nonumber \\
R\,  u_{\phi}^{n,l}(r)&:=& -\left[n A_{n,l}\frac{J_n(\kappa'_{n,l}r)}r
+ B_{n,l} \frac{{\rm d} J_n(\kappa_{n,l}r)} {{\rm d} r} 
\right]  \; .
\label{Normdef3}
\end{eqnarray}

Zeroes in $u_r^{n,l}(r)$ and $u_\phi^{n,l}(r)$ give rise to 
radial node lines of $u_r(r,\phi)$ and $u_\phi(r,\phi)$. A given value
of $l$ does not imply a given number of radial node lines of the
radial or tangential displacement, because the
boundary conditions mix the two displacement components.
The displacements for a few eigenmodes are sketched in Figure \ref{wavevgl}. 
and the frequencies, $\omega_{n,l}$ of the modes
are plotted versus the number of radial nodes $n$ in Figure \ref{Modenplot}.
For collisions, an important characteristic of a mode is the maximal
radial and tangential displacement at the edge of the disk
\begin{equation}
C_{n,l}:=\, u_r^{n,l}(R) \; , \qquad S_{n,l}:=\, u_{\phi}^{n,l}(R) \; .
\end{equation}
The modes can roughly be divided into two classes with
primarily radial or tangential displacement.
In Figure \ref{Modenplot}  radial oscillations
are characterized by  $C_{n,l}>S_{n,l}$, tangential oscillations
by $C_{n,l}<S_{n,l}$.

For $n>2$ two distinct regimes can be distinguished.  For small $l$
the solutions are regularly spaced and the arrangement barely changes,
when going from $n\to n+1$.  The difference in magnitude between
tangential and radial displacement is small (e.g. $C_{n,l}\simeq
S_{n,l}$), and radial as well as tangential modes tend to cluster, the
more so the larger $n$.  For $l\gg n$ the solutions, $\omega_{n,l}$
for the radial and tangential modes appear to be almost independent of
each other, as they are in the case $n=0$. This results in an
irregular spacing of the frequencies of the modes. The distinction
between primarily radial and tangential displacement is pronounced.  Ratios
$C_{n,l}:S_{n,l}$ typically are larger than $1:10$ (or $10:1$) when
$l\gg n$, exceeding $1:100$ for very large $l$.  When two solutions
have very similar frequencies the ratios are smaller at the surface,
but this does not change the overall character of the mode.  The
number of node lines increases uniformly with l in both cases.
As a general statement we conclude, that the mixing of the
displacements introduced by the boundary conditions becomes more
important with increasing $n$ and decreasing $l$.

For $n=1$ dilatation and rotation, stress and 
shear disappear linearily at the origin, but there is a finite displacement
of the center (see Figure \ref{wavevgl}). For $n\ge 2$  displacement of
the center disappears like $r^{n-1}$, dilatation and rotation
like $r^n$, stress and shear go like $r^{n-2}$ 
(and are maximal at the origin for $n=2$). Thus for
increasing $n$ the deformations are increasingly
concentrated towards the edge of the disk.

The mode with the lowest frequency is the quadrupolar mode ($n=2,l=0$),
where contraction in one direction is met by expansion
in the orthogonal direction (see Figure \ref{wavevgl}). 
This mode will turn out to be
the most important mode for the storage of elastic energy,
as suggested by Rayleigh for three-dimensional spheres \cite{Rayleigh}.

\subsection{Energy of vibrations}

We  have a complete orthonormal system of eigenfunctions 
such that
any (symmetric) state of the disk can be expanded in this set:
\begin{eqnarray}
\left(\begin{array}{c} u_r(r,\phi)\\ u_{\phi}(r,\phi)
\end{array}\right)\;=\; \sum_{n,l} Q_{n,l} 
\left(\begin{array}{c} u_r^{n,l}(r) \cos{(n\phi)}\\ 
u_{\phi}^{n,l}(r) \sin{(n\phi)}\end{array}\right).
\end{eqnarray}
To express the elastic energy $E_{\rm elas}$ in terms of the expansion
coefficients,
$Q_{n,l}$, we first use partial integration together with force free
boundary conditions to rewrite the elastic energy as 
\begin{equation}
E_{\rm elas}=-\frac{E}{2(1-\nu^2)}\int_S d^2x \sum_k u_k({\bf x})
({\cal L}{\bf u})_k({\bf x})
\end{equation}
Next, we use the equation for the eigenfunctions of ${\cal L}$
\begin{equation}
({\cal L}{\bf u}^{n,l})_k({\bf x})=-\kappa_{n,l}^2 u_k^{n,l}({\bf x})
\end{equation}
to rewrite the elastic energy as
\begin{equation}
E_{\rm elas}=\frac{m}{2}\sum_{n,l}\omega_{n,l}^2 Q_{n,l}^2.
\end{equation}

The kinetic energy of a  vibrating disk is given by
\begin{eqnarray}
  \label{Kinenerg}
E_{\rm kin}&=&
\frac \rho 2
  \int\limits_0^{2\pi}{\rm d}\phi \int\limits_0^R  {\rm d}r\,r 
(\dot u_r^2(r,\phi)+\dot u_{\phi}^2(r,\phi))  \\
&=& \frac {m}{2}\sum_{n,l} \dot Q_{n,l}^2 \; ,
\end{eqnarray}
so that the total energy of a vibrating disk not interacting with
external forces is given by
\begin{eqnarray}
  \label{Hamiltonfree}
  H^0 := \sum_{n,l}^\infty \Big 
(\frac 1{2m} P_{n,l}^2 + \frac m{2} \omega_{n,l}^2
Q_{n,l}^2\Big )\; ,
\end{eqnarray}
where we have introduced canonical momenta $P_{n,l}=m \dot Q_{n,l}$.
In the elastic approximation a vibrating disk behaves
like a set of independent harmonic oscillators with
canonical loci $Q_{n,l}$ and canonical momenta $P_{n,l}$,
obeying Hamilton's equation of motion
\begin{eqnarray}
\frac{{\rm d} Q_{n,l}} {{\rm d} t} = \frac{\partial H^0}
{\partial P_{n,l}} = \frac{P_{n,l}}m \; , 
\qquad \quad -\frac{{\rm d} P_{n,l}} {{\rm d} t}= 
\frac{\partial H^0}{\partial Q_{n,l}}
= m\omega_{n,l}^2Q_{n,l}
\; .\label{Koeffzus}
\end{eqnarray}

\section{Equations of motion in the presence of a wall}
\label{Koeffs}

The head-on collision of two identical elastic disks is equivalent to 
one elastic disk hitting a hard wall, which
itself is not deformed elastically.
We choose a coordinate system such that the wall is located at $x=0$
(see Figure \ref{Expmodel}) and the disk is approaching from the left. The 
interaction between disk and wall is modelled by a potential 
$V(x)= V_0 e^{\alpha x}$, where $x=x(\phi)$ is the distance between
the edge of the deformed disk and the y--axis.
The choice of $V_0$ is arbitrary, because it only affects the position
of the wall, as can be seen from the substitution 
$V_0e^{\alpha x}= V_0'e^{\alpha(x+(\ln k)/\alpha)}$ with $V_0=kV_0'$.
A hard core interaction is recovered in the limit $\alpha \to \infty$. 
We assume that the disk is moving along the $x$-axis in
positive direction. Its center of mass position is denoted by
$x_0(t)$.

We expand an arbitrary, symmetric state of the disk in normal modes
\begin{eqnarray}
\left(\begin{array}{c} u_r(r,\phi,t)\\ u_{\phi}(r,\phi,t)
\end{array}\right)\;=\; \sum_{n,l} Q_{n,l}(t) 
\left(\begin{array}{c} u_r^{n,l}(r) \cos{(n\phi)}\\ 
u_{\phi}^{n,l}(r) \sin{(n\phi)}\end{array}\right)
\end{eqnarray}
with time dependent expansion coefficients $Q_{n,l}(t)$.
Since the wall is only pushing against the edge of the disk
we need to know the location of the boundary. 
The radial and tangential distortion at the edge are given by
\begin{eqnarray}
u_r(R,\phi,t)= \sum_{n,l} Q_{n,l}(t) C_{n,l} \cos n\phi, \qquad
u_{\phi}(R,\phi,t)= \sum_{n,l} Q_{n,l}(t) S_{n,l} \sin n\phi \; .
\end{eqnarray}
The location $\left(\begin{array}{c} x(\phi,t)
\\y(\phi,t)\end{array}\right)$ of an element  
at $\left(\begin{array}{c}x_0(t)+R\cos \phi\\ R\sin\phi\end{array}\right)$
without deformation, is then given by
\begin{eqnarray}
\left(\begin{array}{c}
x(\phi,t) \\ y(\phi,t) \end{array}\right)=
\left(\begin{array}{c}
x_0(t)+(R+u_r(R,\phi))\cos \phi -u_{\phi}(R,\phi)\sin\phi\\ 
(R+u_r(R,\phi))\sin\phi+u_{\phi}(R,\phi)\cos\phi\end{array}\right) \; .
\end{eqnarray}
Its $x$-component enters into the wall potential and is expressed in
terms of normal modes as follows
\begin{eqnarray}
\label{Deformt}
x(\phi,t):=x_0(t)+R \cos\phi + \sum_{n,l} Q_{n,l}(t)(C_{n,l}\cos n\phi \, \cos
\phi -S_{n,l}\sin
n\phi \, \sin\phi) \; .
\end{eqnarray}
The total energy of an elastic disk interacting with a fixed wall is given by
\begin{eqnarray}
\label{Hamiltonint}
H:= \frac 1{2m}p_0^2 +  \sum_{n,l}^\infty \Big ( 
 \frac 1{2m} P_{n,l}^2 + m \omega_{n,l}^2
Q_{n,l}^2\Big ) 
{}+ V_0 \int\limits_{-\pi/2}^{\pi/2} \hbox {d} \phi
\: e^{\alpha x(\phi,t)} \; .
\end{eqnarray}
The dynamic evolution of the expansion coefficients $Q_{n,l}(t)$ follows from 
Hamilton's equations of motion
\begin{eqnarray}
\frac{{\rm d} Q_{n,l}} {{\rm d} t}& =& \frac{\partial   H}
{\partial P_{n,l}} \;\: ,  \nonumber \\ 
-\frac{{\rm d} P_{n,l}} {{\rm d} t} & = & \frac{\partial   H} 
{\partial Q_{n,l}}  \nonumber \\
 \label{Impulsev1}
 & = & m \omega_{n,l}Q_{n,l}+ \alpha V_0 \int\limits_{-\pi/2}^{\pi/2} {\rm d}\phi
(C_{n,l}\cos n\phi \,\cos\phi - S_{n,l} \sin n\phi \, \sin\phi) 
\, e^{\alpha x(\phi,t)}\; \: .
\end{eqnarray}

Instead of using real valued functions $Q_{n,l}(t)$ and $P_{n,l}(t)$, we find
it convenient to introduce a complex function $q_{n,l}(t)=
q_{n,l}^R(t)+iq_{n,l}^I(t)$ such that
\begin{eqnarray}
Q_{n,l}(t) & = & {\rm Re} \Big(q_{n,l}(t) e^{{\rm i}
  \omega_{n,l}t}  \Big)\; , \nonumber \\
P_{n,l}(t) & = & m \omega_{n,l} {\rm Im} \Big(q_{n,l}(t) e^{{\rm i}
  \omega_{n,l}t}  \Big) \; . \label{compact}
\end{eqnarray}
Hamilton's equation of motion then read
\begin{eqnarray}
\dot q_{n,l}^R \cos{(\omega_{n,l}t)}-\dot q_{n,l}^I
\sin{(\omega_{n,l}t)} & = & 0\; , \nonumber \\
\dot q_{n,l}^R \sin{(\omega_{n,l}t)}+\dot q_{n,l}^I
\cos{(\omega_{n,l}t)} & = & -\frac{\alpha V_0}{m \omega_{n,l}} 
\int\limits_{-\pi/2}^{\pi/2} {\rm d}\phi
(C_{n,l}\cos n\phi \,\cos\phi - S_{n,l} \sin n\phi \, \sin\phi) 
\, e^{\alpha x(\phi,t)} \; .\nonumber
\end{eqnarray}
These two equations can be combined into a single equation for the complex
function $q_{n,l}(t)$
\begin{eqnarray}
  \label{Evolgl}
  \frac{\rm d}  {{\rm d} t} q_{n,l}=
-\frac{{\rm i}\, \alpha V_0}{m\omega_{n,l}} 
e^{-{\rm i } \omega_{n,l} t }\int\limits_{-\pi/2}^{\pi/2} {\rm d}\phi
(C_{n,l}\cos n\phi \,\cos\phi - S_{n,l} \sin n\phi \, \sin\phi) 
\cdot e^{\alpha x(\phi,t)} \; . 
\end{eqnarray}

The time evolution of the center of mass velocity also follows from
Hamilton's equation of motion 
\begin{eqnarray}
  \label{Gesev}
  \frac{{\rm d}   v} {{\rm d} t} = \frac 1m \frac{{\rm d} p_0} 
{{\rm d} t} = -\frac 1m \frac{\partial }{\partial x_0}H = -  \frac
{\alpha V_0}m
\int\limits_{-\pi/2}^{\pi/2} {\rm d}\phi e^{\alpha x(\phi,t)} 
\; .
\end{eqnarray}
completing our description of a two-dimensional
elastic disk hitting a wall.  The only values needed for every
mode $(n,l)$ are the frequency $\omega_{n,l}$ and the maximal
displacements $C_{n,l}$ and $S_{n,l}$.
The actual numerical integration of the equation of evolution
(\ref{Evolgl})
requires the consideration of  many different modes to find a
good estimate of the coefficient of restitution.

\section{Numerical Results}
\label{Numerik}

The actual integration of (\ref{Evolgl}) is numerically demanding,
when high accuracy is required. Time steps have to be set quite low to
capture the oscillations of the fastest modes. Numerical accuracy is
limited by the number of modes that can be used (usually less than
1000), so that a 4th order Runge-Kutta method suffices for the
numerical integration of the equations of motion.  Starting from
$x(\phi,t)$ and $q_{n,l}(t)$, we update $x_0$ according to
Eq.~(\ref{Gesev}) and all $q_{n,l}$ for a given set $(n\leq
n_{\rm max},l\leq l_{\rm max})$ according to Eq.~(\ref{Evolgl}). Then a new
$x(\phi,t+\Delta t)$ is calculated from Eq.~(\ref{Deformt}).  For a
normal collision of a disk with no excited modes the integration over
the edge of the circle can be limited to the upper quadrant because of
symmetry. We typically discretize the upper quadrant with $256$ values
for $\phi$ and retrieve the values of $\cos n\phi$ and $\sin n\phi$
from a look-up-table to effectively speed up the computation. Initial
conditions are $q_{n,l}=0$ for all $n,l$, and a value of $x_0$
sufficiently negative, such that there is no interaction with the
wall.

In our simulations and plots we measure velocity in units
of $c=\sqrt{E/\rho}$, distances in units of $R$ and forces
in units of $mc^2/R=\pi RE$.
The numerical simulations have been performed for a range 
of initial velocities $0.005c \leq v_{\rm i} \leq 0.3c$ and a
a value of $\nu=0.33$ for Poisson's ratio. A few additional runs were made
with other values of $\nu$ to check the dependence
of our results on Poisson's ratio.  
In Figure \ref{Stoss} the collsion of an elastic disk with a rigid
wall is shown, as calculated numerically according to the above procedure.

Computation of the coefficient of restitution requires two different
extrapolations. Modelling disks as elastic continua requires
the limit of infinitely many modes.  At the same time the
artificial exponential potential, introduced to make the collision
accessible to numerical calculations, should be made infinitely hard.

A harder potential of the wall increases the energy, which remains in
the elastic modes for two reasons: 1) The force which excites the
elastic modes is stronger. 2) During the collision less energy is
stored in potential energy, which after the collision is returned to
kinetic energy and hence cannot be transferrred to the internal
degrees of freedom. Raising the number of modes has the opposite
effect. The coefficient of restitution $\epsilon$ is closer to one for
a larger number of modes, because the larger number of modes allows a
better adjustment of the shape of the disk, so that collisions become
effectively ``softer''.  In both cases extrapolation from finite
values is possible, it is more difficult for the number of modes than
for the potential parameter $\alpha$.

To select the right set of modes, we use the following procedure.
For every initial velocity $v_{\rm i}$ the modes 
were ranked according to their 
average energy content in a preliminary run.  
As expected, the average energy content of a mode decreases with
increasing frequency $\omega_{n,l}$.
It turns out that the purely radial modes (see Eq.~(\ref{Radialvib}))
are particularly important. Typically 75 \% of the most 
significant modes have $n=0$. Another essential class of
modes has $l=0$. The importance of these two classes can be inferred from
Figure \ref{Histogramm}, where we show the distribution of energy over
modes (n,l) at closest appraoch.
In Table 1 we list the energy averaged over the entire collision
as well as the energy after collision for the ten most important
modes. It can be seen that the average energy content of a mode
during the collision can be very different from the energy remaining
in the oscillation after the collision has been completed.

\begin{table}
\begin{minipage}[b]{10cm}
\begin{tabular}[ht]{c|cc|cc|}
rank & n &  l  &   average energy &  energy after collision \\
\hline\hline
1 & 2 &  0    &    0.1594843 &   0.0513636 \\ \hline
2 & 3 &  0    &    0.0480930 &   0.0234695 \\ \hline
3 & 0 &  0    &    0.0231257 &   0.0089346 \\\hline
4 & 4 &  0    &    0.0220028 &   0.0038760 \\\hline
5 & 0 &  1    &    0.0128242 &   0.0003144 \\\hline
6 & 5 &  0    &    0.0114396 &   0.0027244 \\\hline
7 & 1 &  1    &    0.0110218 &   0.0002610 \\\hline
8 & 1 &  0    &    0.0066413 &   0.0002548 \\\hline
9 & 6 &  0    &    0.0059540 &   0.0021365 \\\hline
10&  0&  2    &    0.0048560 &   0.0000279 \\
\hline
\end{tabular}
\end{minipage}
\caption[Ranking]
{\footnotesize The ten most important modes ranked according to the
average fraction of the total energy during collision
for $\alpha=500R, v_{\rm i}=0.1c$ using 1600 modes. }

\end{table}

\
In Figure \ref{PotentialSkal} the energy which is absorbed in
the elastic modes is plotted versus potential parameter $\alpha$.
It scales nicely with $1/\alpha$, because the energy that is
stored in the wall potential at closest approach 
is also $\propto 1/\alpha$. Hence the extrapolation $\alpha \to
\infty$ is straightforward, except for small initial velocity, when
 the coefficient of
restitution for a given
potential parameter $\alpha$ approaches one. To identify the inelastic 
signature of the collisions one has
to choose ever harder potentials. This makes it difficult to
extrapolate to the limit $v_{\rm i} \to 0$ numerically. Fortunately the
quasistatic limit can be treated analytically, as discussed in 
the next section.

The extrapolation of the energy stored in vibrations to infinitely many modes
is shown in Figure \ref{Modenskal}. The limit $N \to \infty$ is 
approached like $1/\sqrt{N}$.  The more violent collisions 
need a higher number of modes for the $1/\sqrt{N}$ regime to appear.

The two extrapolation procedures have been performed for a range of
initial velocities and two values of Poisson's number. The resulting
coefficient of restitution $\epsilon$ is
plotted versus the initial velocity in Figure \ref{Zweivgl}.
It is seen to approach one in the quasielastic limit and to decrease
with increasing relative velocity. Collisions are found to be more
inelastic for higher values of $\nu$.

\section{Quasistatic approach}
\label{Static}

As we just have seen, the limiting case $v_{\rm i}/c \ll 1$
is difficult to treat numerically.
However Hertz' law of contact can be extended to two dimensions.
Solving the problem in two dimensions is
actually more complicated than the
three-dimensional case, because there are no local solutions.

The quasistatic acceleration of a disk
by a hard wall is equivalent to a disk in equilibrium 
in a gravitational field and supported by a hard wall. This problem
has been solved for a point contact by H.~Michell
\cite{Punktdisk}. To compute the compression caused by the
gravitational field we have to generalize his solution to an extended contact
between the disk and the supporting wall.

In \cite{contactmech}  the contact pressure
is calculated for a two-dimensional contact of cylindrical
bodies. This solution can be taken over to our problem of a
two-dimensional disk with one-dimensional loading. 
Inside the loaded region $-a \leq y \leq a$  (see Figure \ref{Quasstat})
a normal stress $p(y)$ given by 
\begin{equation}
  \label{Kraftvert}
  p(y)= \frac{2P}{\pi a^2}(a^2-y^2)^{1/2}   
\end{equation}
acts on the surface.
Here $P$ is the total load, $R$ is the radius of the disk and
$a^2=4PR/(\pi E)$. 
Integrating over the loaded region yields 
the stresses in the elastic disk due to the normal pressure $p(y)$:
\begin{eqnarray}
  \label{Bestgleichgn}
  \sigma_{yy}(x,y)&=& \frac{-2x}{\pi}\int\limits_{-a}^a \frac{p(s)(y-s)^2
  \hbox{d}s} {[(y-s)^2+x^2]^2} \; ,\nonumber \\
 \sigma_{xx}(x,y)&=&-\frac{2x^3}{\pi}  
\int\limits_{-a}^a \frac{p(s)
  \hbox{d}s} {[(y-s)^2+x^2]^2} \; ,\nonumber \\ 
\sigma_{xy}&=&-\frac{2x^2}{\pi}  
\int\limits_{-a}^a \frac{p(s) (y-s)
  \hbox{d}s} {[(y-s)^2+x^2]^2} \; .
 \end{eqnarray}
Along the axis of symmetry integration is possible and one finds
\begin{eqnarray}
  \label{Spannentlangz}
  \sigma_{yy}(x,0) &=& -\frac{2P}{\pi a^2} \: \Big
  (\frac{a^2+2x^2}{(a^2+x^2)^{1/2}}-2x \Big) \; ,\\
  \sigma_{xx}(x,0) &=& -\frac{2P}{\pi}(a^2+x^2)^{-1/2}\; , \nonumber
\qquad \sigma_{xy}(x,0) \; =\; 0 \; .
\end{eqnarray}

In a three-dimensional elastic medium, the displacement decreases
like $1/r$ for large distances $r$ from the loaded region. In a
two-dimensional elastic medium with one-dimensional loading the
displacement varies as $\ln (r)$, so that the displacement can only
be defined relative to an arbitrarily chosen datum (which for
three-dimensional systems is usually taken at infinity). This implies
that the total deformation $\delta$ cannot be computed from the local
stress distribution $p(y)$.
To find the total compression for a two-dimensional system
one has to consider the stress distribution in the bulk of the
two-dimensional elastic body as well as its shape and size.  

To this end we need to generalize the solution of Michell
\cite{Punktdisk}, who solves the problem of a heavy disk supported by
a point force. The stress generated by the gravitational force is
most easily evaluated in a coordinate system whose origin is located
in the center of the sphere. The equations of equilibrium for a
uniform downward force of magnitude $\rho g$ read
\begin{eqnarray}
  \label{Gravtest}
  \frac{\partial \sigma_{xx}}{\partial x}  +
 \frac{\partial \sigma_{xy}}{\partial y} = \rho g\; , \qquad
 \frac{\partial \sigma_{yy}}{\partial y} +  
\frac{\partial \sigma_{xy}}{\partial x} = 0\; , \qquad
\nabla^2(\sigma_{yy}+\sigma_{xx})=0 
\end{eqnarray}
and are solved by
\begin{eqnarray}
  \label{Schwerkraft}
  \sigma_{xx} = \frac{1}{2}\rho g x\;,  \qquad  \sigma_{yy}= -\frac{1}{2}
\rho g x \; ,\qquad \sigma_{xy}= \frac{1}{2}\rho g y\; .
\end{eqnarray}
The stress due to the point contact is purely radial in a coordinate
system whose origin lies in the point of contact
\begin{equation}
\sigma_{rr}=-2\rho g R^2 \frac{\cos \phi}{r} \; .
\label{Pointstress}
\end{equation}
To obtain the total stress distribution we transform the stress due to
gravitation (\ref{Schwerkraft}) to the coordinate system whose 
origin lies in the point of contact 
\begin{eqnarray}
  \sigma_{xx} = \frac{1}{2}\rho g (x-R)\;,  \qquad  \sigma_{yy}= -\frac{1}{2}
\rho g (x-R)\; , \qquad \sigma_{xy}= \frac{1}{2}\rho g y .
\end{eqnarray}
and superimpose the two contributions to the stress.
The resulting solution does not satisfy the boundary conditions of
stress free edges, which can be guaranteed by adding a uniform biaxial
tension $\sigma_{xx}=\sigma_{yy}=\rho g R$. Hence the total stress is
given by
\begin{eqnarray}
\sigma_{xx} &=& \rho g \Big( x/2 -\frac{2 R^2 x^3}{(x^2+y^2)^2}\Big)
 \nonumber\; , \\
 \sigma_{yy} &=& -\rho g \Big( (x-2R)/2+\frac{2R^2x y^2}{(x^2+y^2)^2} \Big )
\; . \end{eqnarray}

The radial stress distribution (\ref{Pointstress}) 
from the point contact (which creates a
logarithmic divergence in the displacement) will now be replaced 
by the realistic stress distribution(\ref{Spannentlangz})  for an
extended contact.
The total stress along the axis of symmetry $y=0$ is then given by
\begin{eqnarray} \nonumber
 \sigma_{xx}(x,0) &=& -\frac{P}{\pi} \Big(\frac{2}{(a^2+x^2)^{1/2}} 
 - \frac{x}{2R^2} \Big ) \; , \\
\sigma_{yy}(x,0) &=& -\frac{P}{\pi } \: \Big
  (\frac{2(a^2+2x^2)}{a^2(a^2+x^2)^{1/2}}-\frac{4x}{a^2}+\frac{x-2R}{2R^2}
\Big) \; .
\end{eqnarray}
Here the total load is simply given by $P=m g= \pi R^2 \rho g$.

The strain can be calculated from the formula
\begin{equation}
\epsilon_{xx}=  \frac 1{E} \big(\sigma_{xx} - \nu \sigma_{yy}
\big)
\end{equation}
and the total compression $\delta$ of the disk is found by integrating 
$\epsilon_{xx}$ from $x=0$ to $x=2R$. For $a\ll R$ we find
\begin{eqnarray}
\delta = -\int\limits_0^{2R} \epsilon_{xx} \hbox{d}x &=&
\frac{P}{\pi E} (2 \ln(4R/a)-1-  \nu) \\
&=&
\label{Kraftvollst}
 \frac {P}{\pi E} \: 
\Big(\ln \big(\frac{4 R \pi E}P\big)-1-\nu \Big) \; .
\end{eqnarray}

The compression $\delta$ can also be obtained from the related problem
of a disk compressed between two walls, which has been solved 
in \cite{contactmech2}. In this case the total deformation 
is twice the deformation as calculated above for
small $\delta$.

In Figure \ref{Statplot} we show $P(\delta)$ as computed by three
different methods. The analytical
results are compared to the dynamic calculations of Section
\ref{Numerik}.  If the disk is held fixed at a given distance from the
wall, it will assume a deformation which minimizes its total energy.
This deformation and the force acting on the disk can be computed
using the expansion of the elastic deformation by the modes calculated
in section \ref{Moden}. The total energy given by
Eq.~(\ref{Hamiltonint}) with $P_{n,l}=0$ is iteratively minimized
until convergence is achieved.  The numerical result of this
calculation is shown as a dashed-dotted line in Figure \ref{Statplot}.
The results from the quasistatic calculation and from the minimization
procedure agree well for small $\delta$, whereas for larger
deformations one observes deviations from Hertz law of contact.  The
forces during the dynamical collision are considerably higher than in
the quasistatic limit, because the compression is localized.  The
results of the dynamic calculation are shown for two initial
velocities to demonstrate the approach towards the quasistatic case
for the smaller initial velocity.

In the limit of small $P$, Eq.~(\ref{Kraftvollst}) reads
\begin{equation}
\label{deltavonP}
\delta = \frac P {\pi E} \ln\big(\frac{4 R \pi E} P \big ) \; ,
\end{equation}
and may be
inverted to yield
\begin{eqnarray}
P=  \frac {\delta \pi E}{\ln {4R}/ \delta} \;+\; 
{\cal O}\big( \frac {\delta \ln \ln 1/\delta}{(\ln \delta)^2}  \big)
\quad .
\end{eqnarray}
To lowest nontrivial order the potential energy is given by
\begin{eqnarray}
V\simeq \frac 12 \delta^2 \frac{\pi E}{\ln (4R/\delta)} \quad .
\end{eqnarray}
The kinetic energy before collision is given by 
$\frac 12 v_{\rm i}^2 \pi R^2 \rho$. In the limit of low $v_{\rm i}$, 
where the
quasistatic approximation is expected to hold, conservation of energy
can be used to obtain the maximum value of $\delta$
\begin{eqnarray}
\delta_{\rm max} \simeq \frac {v_{\rm i}} c R \sqrt{\ln \frac {4c}{v_{\rm i}}} 
\qquad
\hbox{and hence} \qquad  \tau \propto \frac Rc\sqrt{\ln \frac{4c}{v_{\rm i}}} 
\label{xmaxapprox}
\end{eqnarray}
for the contact time $\tau$.  In the limit $v_{\rm i} \to 0$ the
contact time diverges logarithmically, i.e. much slower than for
three-dimensional spheres, where Hertz' theory predicts $\tau \propto
v_{\rm i}^{-1/5}$.  In Figure \ref{Contacttime} contact time and
maximal compression are plotted with the force calculated from
Eq.~(\ref{Kraftvollst}) as a function of the initial velocities.  The
very slow divergence of the contact time with $v_{\rm i} \to 0$ can be
observed. The results for the contact time as obtained from the
quasistatic calculation agree well with the numerical simulations of
the full dynamic problem in the range of velocities were both methods
apply.

Our aim is an approximate expression for the 
coefficient of restitution $\epsilon$ in the limit of small
velocities $v_{\rm i}$. This can be achieved by  using
the quasistatic force law in the equations of motion.
Our approach is a generalization
of Rayleigh's \cite{Rayleigh}, who derived
the energy stored in the fundamental mode in the case of
spheres.

We replace the wall potential in
(\ref{Hamiltonint}) by a potential $V$, that acts only
 at the point $\phi=0$ of the boundary.  This amounts to
\begin{eqnarray}
\label{Hamilstat}
H:= \frac 1{2m}p_0^2 +  \sum_{n,l}^\infty \Big ( 
 \frac 1{2m} P_{n,l}^2 + m \omega_{n,l}^2
Q_{n,l}^2\Big ) 
+ V(x(\phi=0,t))\; , \\
\hbox{where } \qquad x(\phi=0,t)= x_0(t)+R +\sum_{n,l}
 Q_{n,l}(t)C_{n,l} \; . \qquad 
\end{eqnarray}
From Hamilton's equations of motion we derive
\begin{eqnarray}
-\frac{{\rm d} P_{n,l}} {{\rm d} t} & = & \frac{\partial   H} 
{\partial Q_{n,l}}  
 \; =\;  m \omega_{n,l}Q_{n,l}+ C_{n,l} \frac{\partial V}{\partial x} \; . 
\end{eqnarray}
Next we approximate $\partial V/\partial x$ by the force $P$ as  calculated
from (\ref{Kraftvollst}).  Using the compact notation (\ref{compact})
we can write
\begin{eqnarray}
\frac{\rm d} {{\rm d} t}\, q_{n,l} &=& - {\rm i} e^{-{\rm i} \omega_{n,l}t}
\frac{C_{n,l}}{m \omega_{n,l}} P(t)\; , \\
\frac{\rm d}  {{\rm d} t}\,v&=&-\frac 1m P(t)\; , \\
\frac{\rm d} {{\rm d} t}   \,\delta &=& -v
\quad .
\end{eqnarray}
These equations can be integrated numerically with initial conditions
$q_{n,l}=0,\delta=0,v=v_{\rm i}$. The energy which
remains in the modes after collision is given by
\begin{eqnarray}
H_{\rm ex}= \frac m2 \sum_{n,l} \omega_{n,l} |q_{n,l}(\tau)|^2 \; .
\end{eqnarray}

In Figure \ref{ContacEnergy} we show  how much energy
is stored in the modes for low $v_{\rm i}$. The quasistatic approximation
agrees in magnitude with the result from the dynamical calculation.
Most of the energy goes into the quadrupolar mode. Since the
quasistatic approach is only correct to first order, 
the decreasing vibrational energy for $v_{\rm i}>10^{-2}c$ 
is presumably unphysical.

\section{Conclusions}
It was our aim to derive the coefficient of restitution as a function
of velocity, starting from a microscopic model of simple objects with
internal degrees of freedom, which can absorb part of the kinetic
energy of translation. We have discussed in detail the head-on collision of
two elastic disks, initially nonvibrating. This problem is equivalent
to the collision of an elastic disk with a rigid wall, representing the
plane of reflection symmetry of the two colliding disks.

The dynamics of a collision has been formulated with help of
Hamilton's equations of motion for the normal coordinates of the elastic
disk, interacting with a repulsive wall potential. The resulting
dynamic equations were solved numerically  for a finite number of
modes and then extrapolated to the case of infinitely many modes.

The main results are the following.  In contrast to the quasistatic
theory of Hertz, the dynamic collision of two identical elastic
spheres is inelastic in the sense that the relative velocity is
decreased, corresponding to a coefficient of restitution smaller than
one. The amount of translational energy which is converted to
vibrational energy depends on relative velocity and Poisson's number
$\nu$. The conversion is more effective for materials with low shear
modulus, corresponding to a value of $\nu$ close to one. For high
initial velocities the coefficent of restitution can be rather small,
e.g.  $\epsilon \sim 0.65$ for $\nu=0.9$ and $v_{\rm i}=0.3 c$. In the
quasistatic limit, $v_{\rm i} \to 0$, collisions become more and more elastic.
We have generalized the quasistatic approach of Hertz and Rayleigh to
two dimensions and shown that the contact time $\tau$ diverges
logarithmically as $v_{\rm i} \to 0 $ and the maximum deformation
vanishes as
$\delta_{\rm max} \sim v_{\rm i} \tau$. The dynamic approach agrees with the
quasistatic theory in the regime, where both theories apply.

We expect the collisions to be more strongly inelastic, if the two
spheres have different size. For one-dimensional elastic rods we could
show that the coefficient of restitution is strictly one, if the two
rods have equal length, no matter whether vibrations are excited
before collsion or not. In the latter case, when no vibrations are
excited, the coefficient of restitution is given by the ratio of
lenghts of the two rods. We are presently extending our calculations
to disks of different radii. Future perspectives include collisions of
elastic spheres as well as more elaborate models of contact,
e.g. including roughness.
 
\section*{Acknowledgement}
We thank Timo Aspelmeier, Reiner Kree, and Peter Mueller for
 interesting and helpful discussions.

\section*{Appendix}

In this appendix we show that the differential operator $\cal L$
as defined in (\ref{Bewgleich}) is hermitean with force free
boundary conditions, which read
\begin{eqnarray}
\sigma_{rr}(R,\phi)&=& \left[\frac{\partial u_r}{\partial r} +
 \nu\Big(\frac{u_r}r +\frac 1r
\frac{\partial u_\phi} {\partial \phi}
\Big) \right]_{r=R} = 0
\nonumber\; , \\ \label{boundary}
\sigma_{r\phi}(R,\phi)&=&\left[\frac{\partial }  {\partial u_r}\phi + 
\frac 1r \frac{\partial } {\partial u_\phi}\phi
-\frac{u_\phi}r \right]_{r=R} \quad = 0 \; .
\end{eqnarray}

We have for two arbitray displacement fields $\bf u$ and
$\bf v$
\begin{eqnarray}
\int\limits_{\cal S} {\rm d}x^2\, {\bf v} \cdot \Big (
\frac{1+\nu}{2}\nabla(\nabla\cdot{\bf u})+\frac{1-\nu}{2}\Delta{\bf u}
\Big )
&=& \int\limits_{\cal S}{\rm d}x^2\, {\bf u} \cdot \Big (
\frac{1+\nu}{2}\nabla(\nabla\cdot{\bf v})+\frac{1-\nu}{2}\Delta{\bf v}
\Big ) \nonumber \\ 
&& \;+\quad \hbox{ boundary terms} \; .
\end{eqnarray}
$\cal L$ is hermitian, if the boundary terms vanish with 
 $\bf u$ and $\bf v$ satisfying
Eq.~(\ref{boundary}).
Using the normal vector $\hat {\bf r}$ along the boundary
and d$s = r\,$d$\phi$, the boundary terms  read
\begin{eqnarray}
  \label{Randterme}
&&  \frac{1+\nu}2 \int\limits_{\partial \cal S} {\rm d}s
\Big ( (\nabla \cdot {\bf u}) (\hat{\bf r} \cdot{\bf v}) -
 (\nabla \cdot {\bf v})(\hat{\bf r} \cdot{\bf v}) \Big )
\nonumber \\ &&\quad+ \quad
  \frac{1-\nu}2 \int\limits_{\partial \cal S} {\rm d}s
\Big ( {\bf v}\:(\hat{\bf r} \cdot \nabla ) \cdot{\bf u}) -
{\bf u}\:(\hat{\bf r} \cdot \nabla ) \cdot{\bf v})
 \Big )\\
&=&  \frac{1+\nu}2 \int\limits_{\partial \cal S} {\rm d}s
\Big(v_r (\frac{\partial  u_r}{\partial r} + \frac {u_r}r + \frac 1r
\frac{\partial  u_\phi}{\partial \phi})  
-u_r (\frac{\partial v_r}{\partial r} + \frac {v_r}r + \frac 1r
\frac{\partial v_\phi}{\partial \phi}) 
 \Big) \nonumber \\ &&\quad +\quad 
\frac{1-\nu}2 \int\limits_{\partial \cal S} {\rm d}s
\Big (
v_r \frac{\partial u_r}{\partial r} +v_\phi \frac{\partial u_\phi}{\partial r}
-u_r \frac{\partial v_r} {\partial r}+u_\phi \frac{\partial v_\phi}{\partial r}
\Big) \\
&=& \int\limits_{\partial \cal S} {\rm d}s \Big (
v_r \Big( \frac{\partial u_r}{\partial r} + \nu\Big(\frac{u_r}r +\frac 1r
\frac{\partial u_\phi}{\partial \phi}
\Big)\Big)
- \frac{1-\nu}2 \Big( u_r \frac{\partial v_\phi}{\partial \phi}  + u_\phi
\frac 1r\frac{\partial v_\phi}{\partial \phi}
 \Big)\Big )
  \nonumber \\
&&\quad - \quad \int\limits_{\partial \cal S} {\rm d}s \Big (
u_r \Big( \frac{\partial v_r}{\partial r} + \nu\Big(\frac{v_r}r+ \frac 1r
\frac{\partial v_\phi}{\partial \phi}
\Big)\Big)
-\frac{1-\nu}2 \Big( v_r \frac{\partial u_\phi}{\partial \phi}  + v_\phi
\frac 1r\frac{\partial u_\phi}{\partial \phi}
 \Big)\Big )\; . \label{Arrang}
\end{eqnarray}
In (\ref{Arrang}) we rearranged the terms for our purposes.
The following relationships are derived using partial integration 
\begin{eqnarray}
\int\limits_0^{2\pi} {\rm d}\phi \; v_\phi \frac{\partial u_r}{\partial \phi}
&=& -\int\limits_0^{2\pi} {\rm d}\phi\; u_r \frac{\partial v_\phi}
{\partial \phi} \; + \;
\underbrace{\Big[v_\phi u_r\Big]_0^{2\pi}}_{=0} \; , 
 \nonumber \\
\int\limits_0^{2\pi} {\rm d}\phi \; v_\phi \frac{\partial u_\phi}
{\partial \phi}
&=& -\int\limits_0^{2\pi} {\rm d}\phi \;u_\phi 
\frac{\partial v_\phi}{\partial \phi} \; + \;
\underbrace{\Big[u_\phi v_\phi\Big]_0^{2\pi}}_{=0} \; .
\label{uvRelat}
\end{eqnarray}
Both times the boundary terms vanish because ${\bf u}$ and $\bf v$
are single valued functions of $\phi$. The same relationships hold
with $ \bf u$ and $\bf v$ interchanged.
We use these and subtract a term $u_\phi v_\phi /r$
from both integrals  to finally  write the boundary terms
as
\begin{eqnarray}
\int\limits_{\partial \cal S} {\rm d}s \Big \{
v_r \Big( \frac{\partial u_r}{\partial r} + \nu\Big(\frac{u_r}r +\frac 1r
\frac{\partial u_\phi}{\partial \phi}
\Big)\Big)
+ \frac{1-\nu}2 \: v_\phi \Big(
\frac{\partial u_r}{\partial \phi} + 
\frac 1r \frac{\partial u_\phi}{\partial \phi}
-\frac{u_\phi}r 
 \Big)  \Big \} \quad&& \nonumber \\
 -\:  \int\limits_{\partial \cal S} {\rm d}s \Big \{
u_r \Big( \frac{\partial v_r}{\partial r} + \nu\Big(\frac{v_r}r +\frac 1r
\frac{\partial v_\phi}{\partial \phi}
\Big)\Big)
+ \frac{1-\nu}2 \: u_\phi \Big(
\frac{\partial v_r}{\partial \phi} + 
\frac 1r \frac{\partial v_\phi}{\partial \phi}
-\frac{v_\phi}r 
 \Big) \Big \} & \stackrel != & 0 \qquad
\label{Boundend}
\end{eqnarray}
Obviously, the lefthand side of (\ref{Boundend}) vanishes
if the boundary conditions (\ref{boundary}) are fulfilled
by $\bf u$ and $\bf v$.
Finally we observe that $\cal L$ is also hermitian
with respect to a fixed boundary $u_r=u_\phi = 0$ and
$v_r=v_\phi = 0$.

\newpage
{\Large Captions}

\begin{figure}[htb]\epsfxsize=12cm
\epsfbox{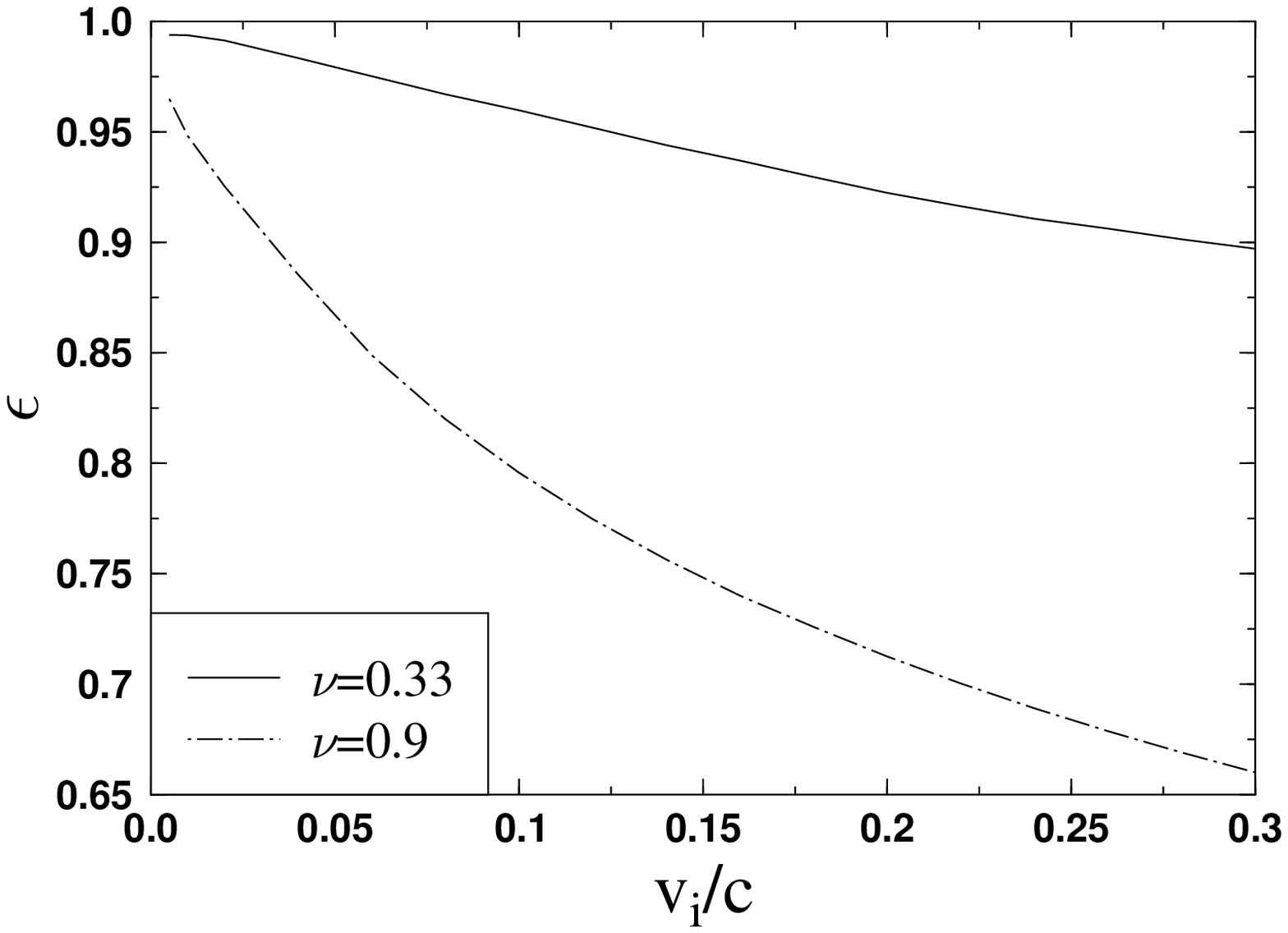}
\caption[Coefficient of restitution]
{}
\label{Zweivgl}
\end{figure}

Figure \ref{Zweivgl}:
Coefficient of restitution $\epsilon$ as a function
of initial velocity for $\nu=0.33$ and $\nu=0.9$.
Here $v_{\rm i}$ denotes the initial relative velocity, expressed in 
units of $c=\sqrt{E/\rho}=1$ (see below).

Figure \ref{wavevgl}:
Plot of the displacement of three simple eigenmodes.
Radii and circles of the originally undeformed disk assume the
distorted shapes sketched here.

\begin{figure}[htb]\epsfxsize=12cm
\epsfbox{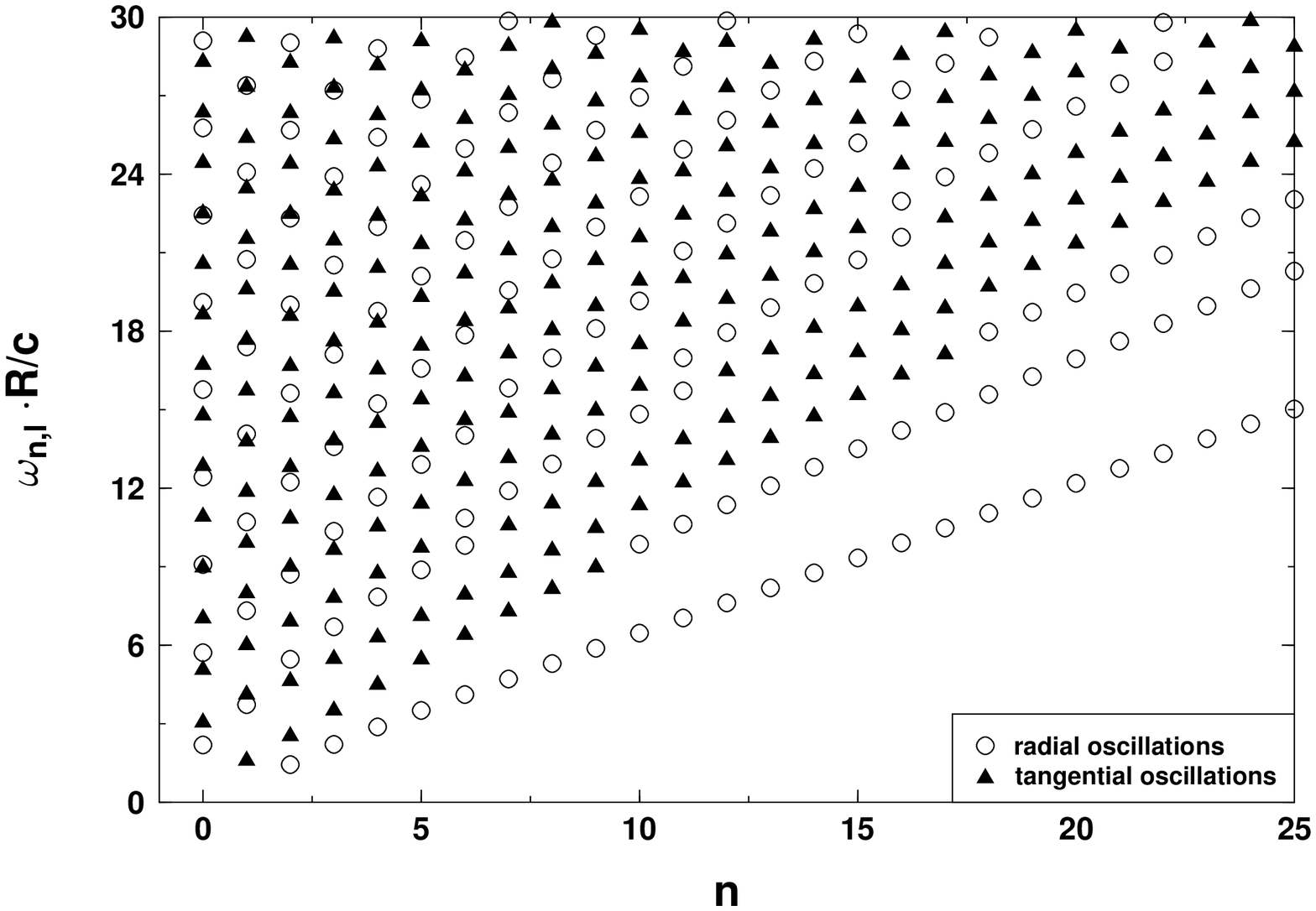}
\caption[Frequency distribution]
{}  
\label{Modenplot}
\end{figure}

Figure \ref{Modenplot}:
Dimensionless eigenfrequencies $\omega_{n,l}\cdot R/c$ 
of an elastic disk versus $n$ 
($\nu=0.33$). We distinguish radial and tangential oscillations,
according to the dominant displacement at the edge of the disk.

\begin{figure}[ht]\epsfxsize=12cm
\epsfbox {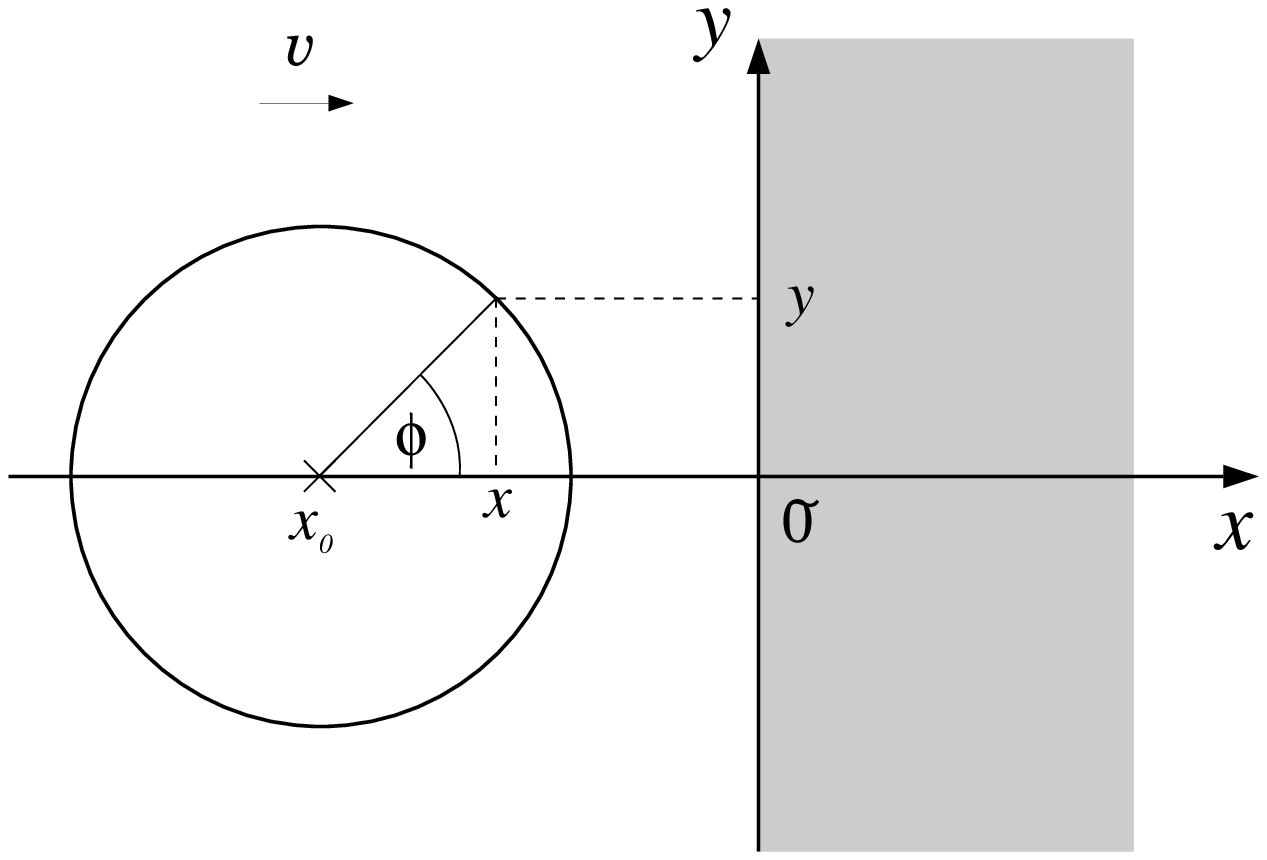}
\caption[Modelling the collision]
{}
\label{Expmodel}
\end{figure}

Figure \ref{Expmodel}:
Illustration of the model of an elastic disk hitting a
rigid wall.

\begin{figure}[ht]\epsfxsize=7cm
\epsfbox {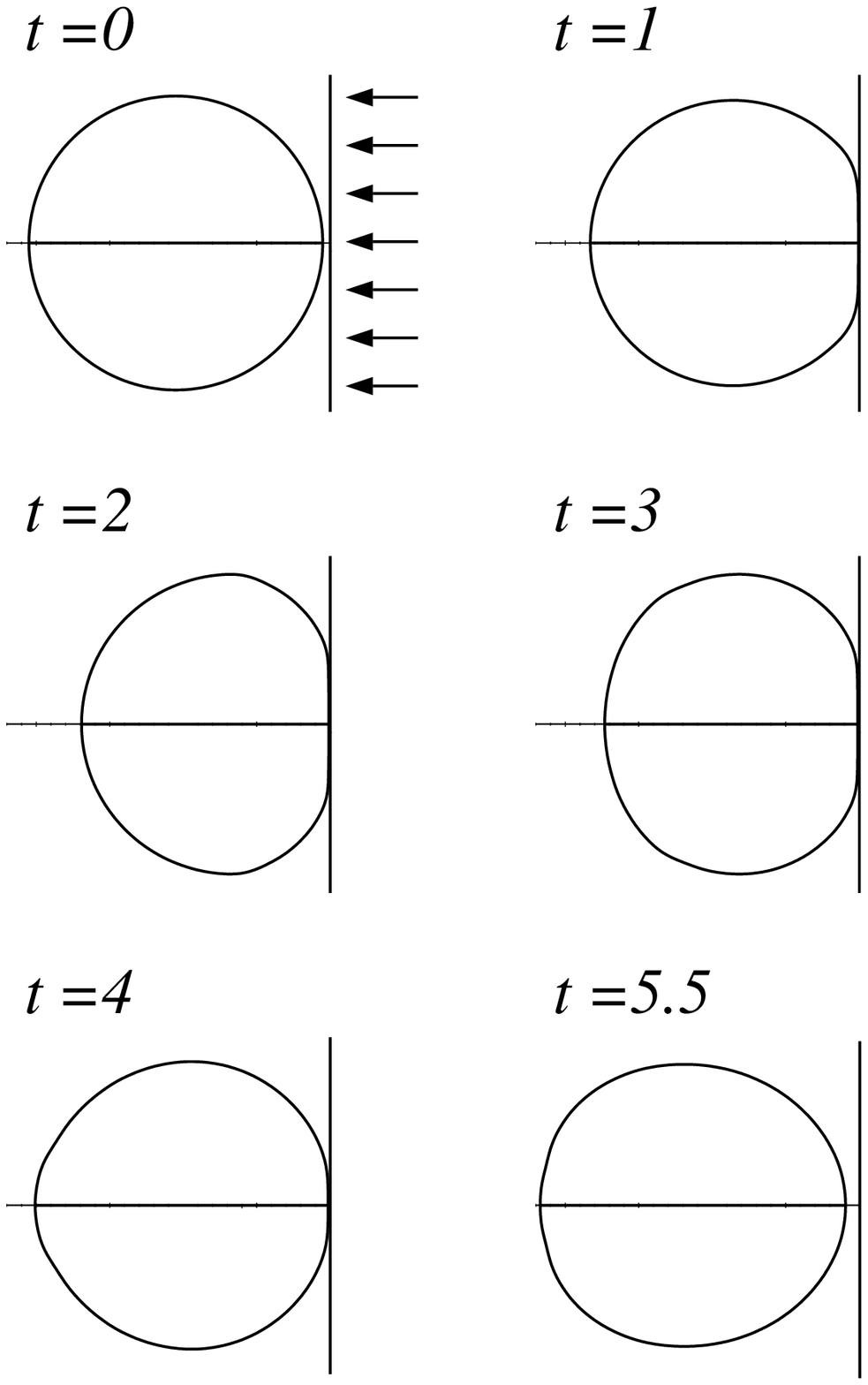}
\caption[Collision of disk with a hard wall]
{}
\label{Stoss}
\end{figure}

Figure \ref{Stoss}:
Time sequence of a disk with $v_{\rm i}=0.2 c$ hitting a rigid wall
located at the origin. The deformations remaining after the
the collision can be clearly identified.

\begin{figure}[ht]\epsfxsize=10cm
\epsfbox{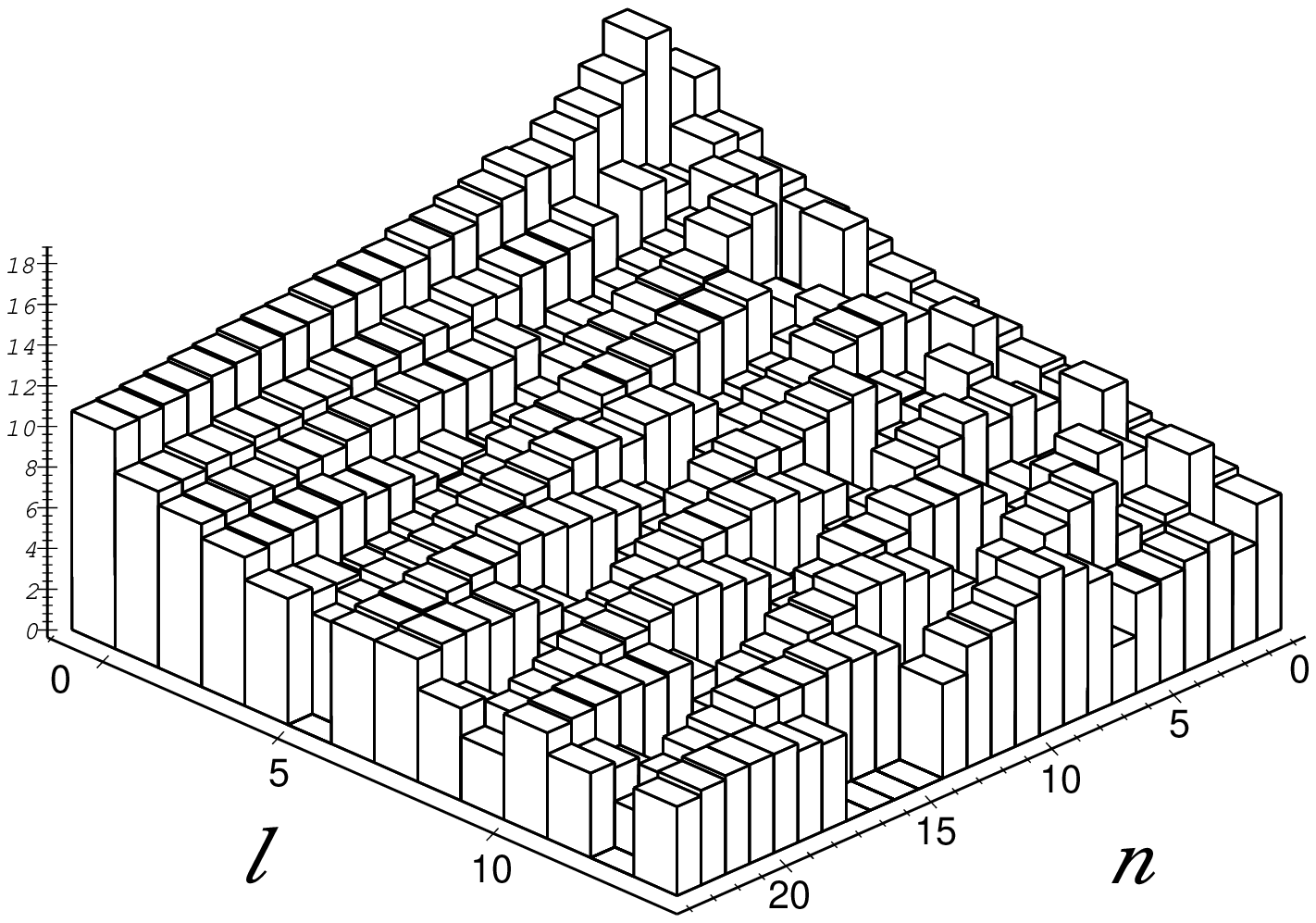}
\caption[Energy spectrum of modes]
{}
\label{Histogramm}
\end{figure}

Figure \ref{Histogramm}:
Logarithm of energy (arbitrary units)
in modes with low $(n,l)$ at closest approach.
The spectrum is dominated by the
quadrupolar mode $(n=2,l=0)$. 

\begin{figure}[htb]\epsfxsize=12cm
\epsfbox{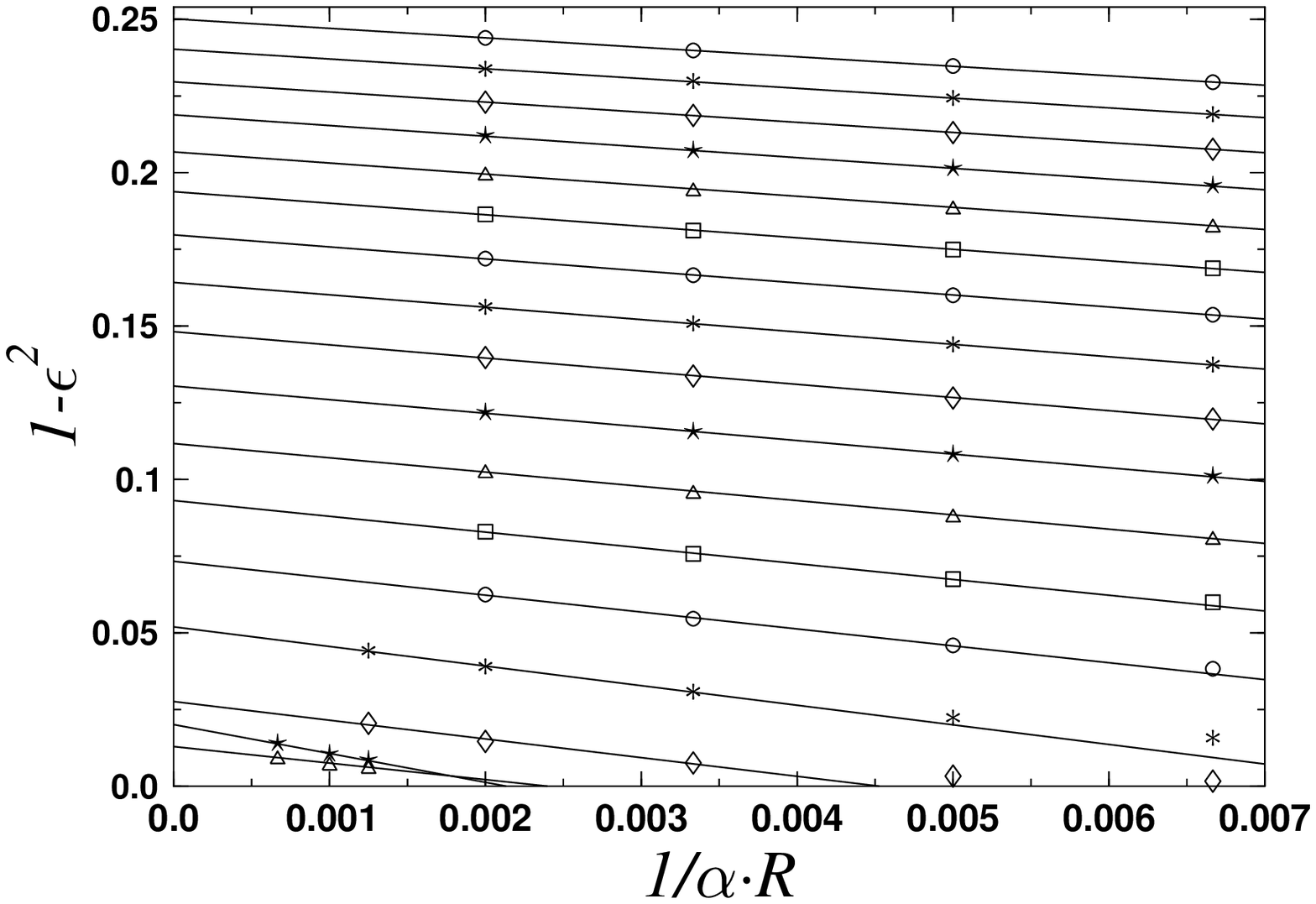}
\caption[Skaling against potential]
{}
\label{PotentialSkal}
\end{figure}

Figure \ref{PotentialSkal}: 
Fraction of the energy stored in the elastic modes $1-\epsilon^2$
after collision with a wall potential $V=V_0 e^{\alpha x}$,
extrapolated to $\alpha \to \infty$. The different curves (from bottom
to top) correspond to different values of initial velocity: $v_{\rm
  i}/c=0.005, 0.01, 0.02, 0.04, 0.06, \ldots, 0.30$.  For $v_{\rm
  i}\le 0.1c$ only the three most significant data points were used to
determine the regression lines. $N=940$ modes were taken into
account.

\begin{figure}[htb]\epsfxsize=12cm
\epsfbox{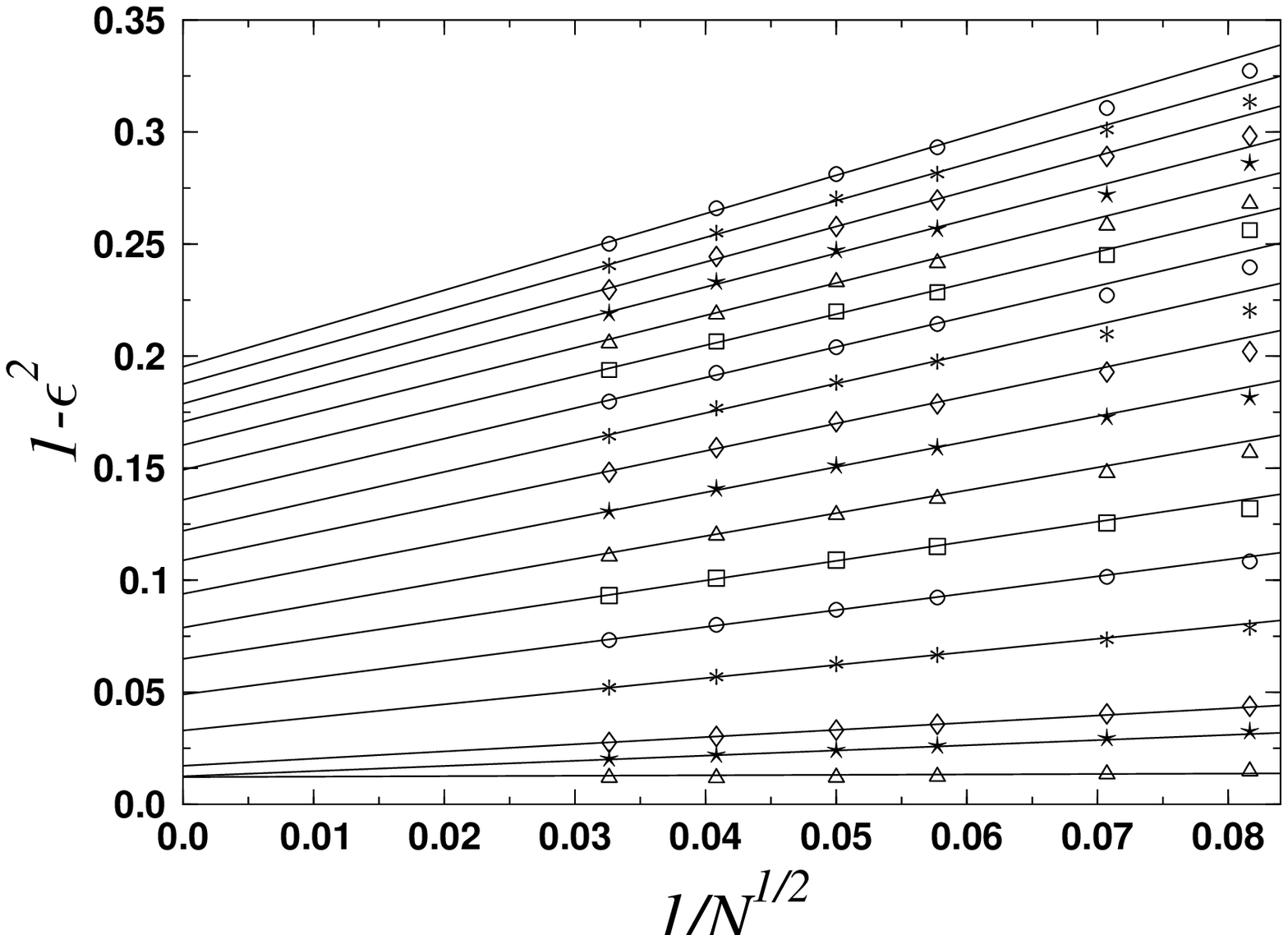}
\caption[calculating the curve]
{}
\label{Modenskal}
\end{figure}

Figure \ref{Modenskal}:
Fraction of energy stored in vibrational modes extrapolated to
$N\to \infty$.  The energy scales with $1/\sqrt{N}$.
For the regression lines $N=300,400,600$ and $940$ were used.

\begin{figure}[htb]\epsfxsize=7cm
\epsfbox{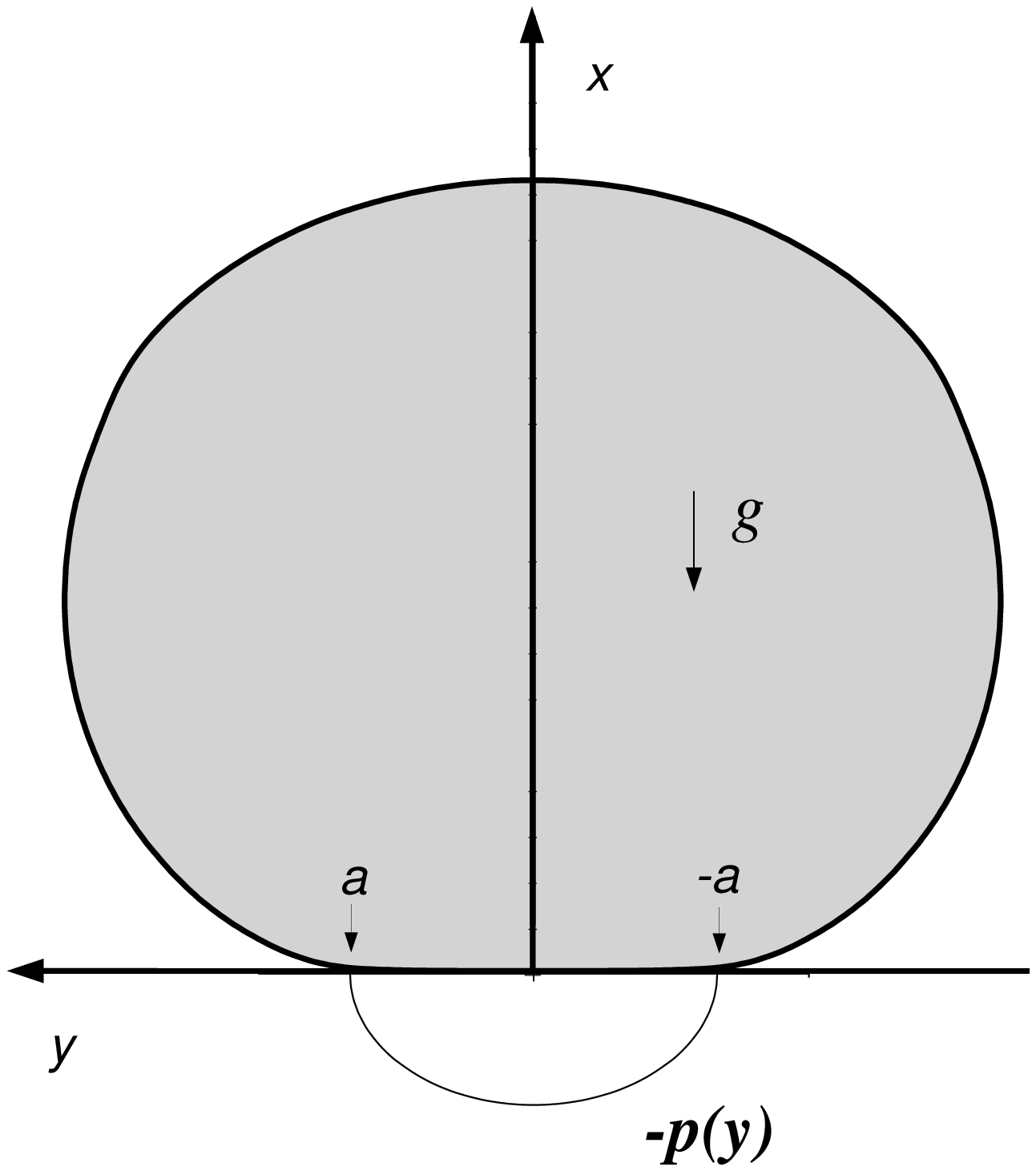}
\caption[Visualization of the quasistatic approach]
{}
\label{Quasstat}
\end{figure}

Figure \ref{Quasstat}:
Visualization of the quasistatic approach.

\begin{figure}[htb]\epsfxsize=12cm
\epsfbox{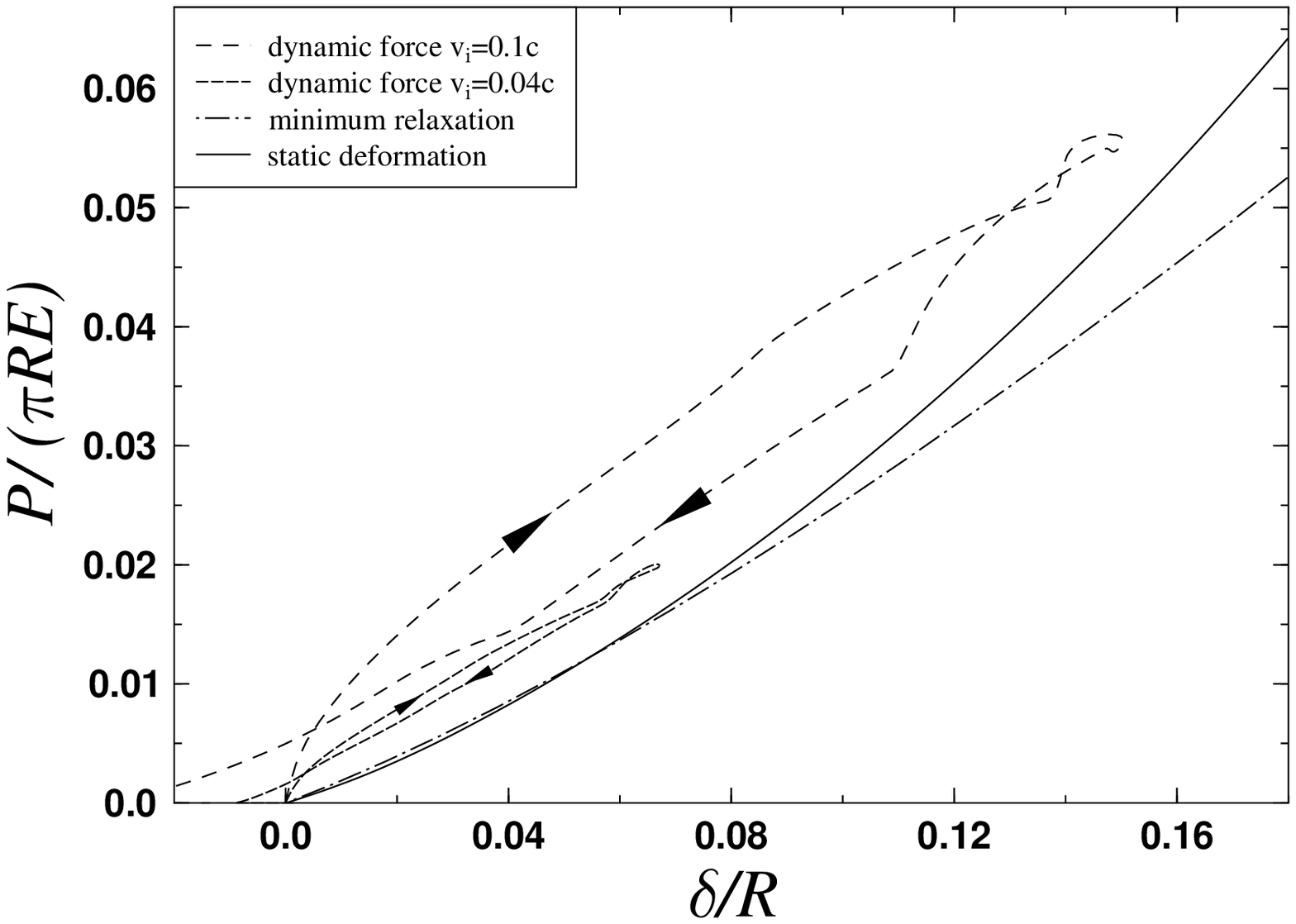}
\caption[Force versus compression]
{}
\label{Statplot}
\end{figure}

Figure \ref{Statplot}:
Dimensionless force versus compression as calculated with different approaches.
For the dynamic calculation we have used $\alpha/R=500$ and two values
of initial relative velocity, $v_{\rm i}/c=0.1$ and $v_{\rm i}/c=0.04$.
The arrows indicate the direction of time.
1300 modes were used for the relaxation towards the minimal energy
for a given distance.

\begin{figure}[htb]\epsfxsize=12cm
\epsfbox{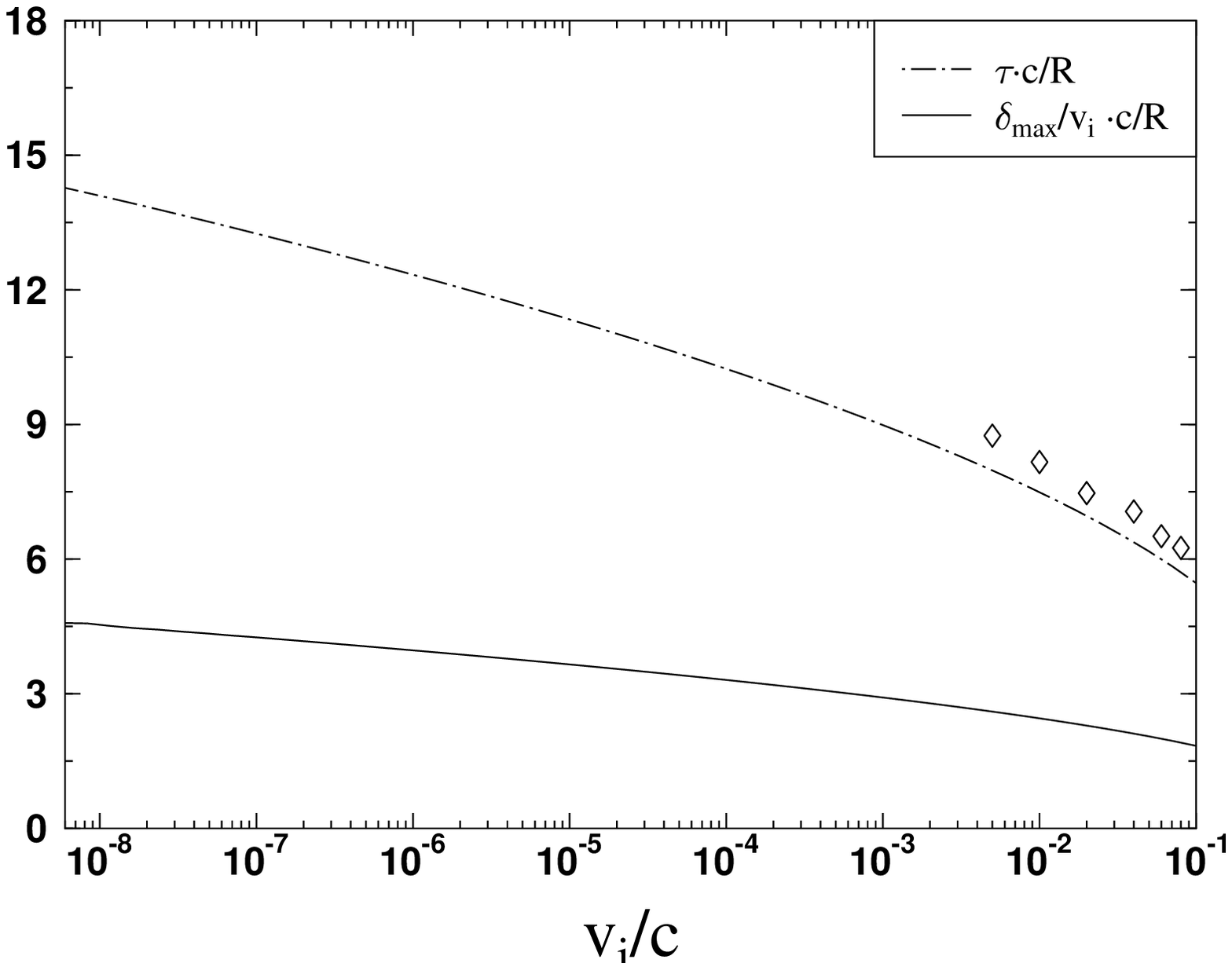}
\caption[Contact time versus initial speed]
{}
\label{Contacttime}
\end{figure}

Figure \ref{Contacttime}:
Dimensionless contact time $\tau \cdot R/c $ and rescaled
maximal compression 
$(\delta_{\rm max}/R) (c/v_i) $ as calculated from
the quasistatic approximation.  The result for $\tau$ compares 
well with  data points ($\Diamond$) from the dynamical calculation.

\begin{figure}[htb]\epsfxsize=12cm
\epsfbox{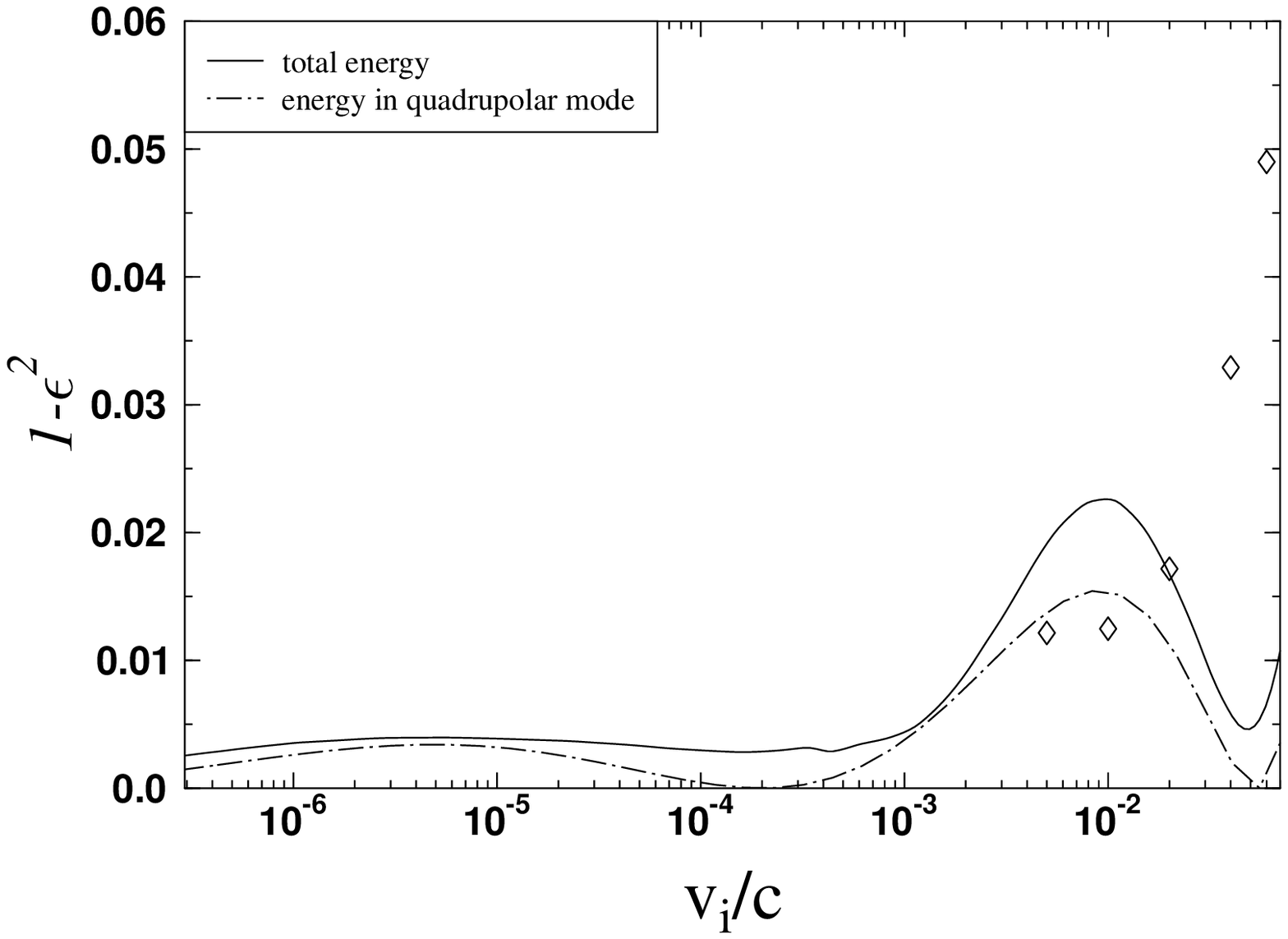}
\caption[Static energy]
{}
\label{ContacEnergy}
\end{figure}

Figure \ref{ContacEnergy}:
Fraction of energy stored in the lowest 1800 modes
versus velocity using the Rayleigh 
approach.  For these very low velocites the 
energy uptake is dominated by the quadrupolar mode (n=2,l=0). For
comparison we show data points ($\Diamond$) from the full dynamic
calculation.

\newpage

\begin{figure}[htb]\epsfxsize=6cm
\epsfbox{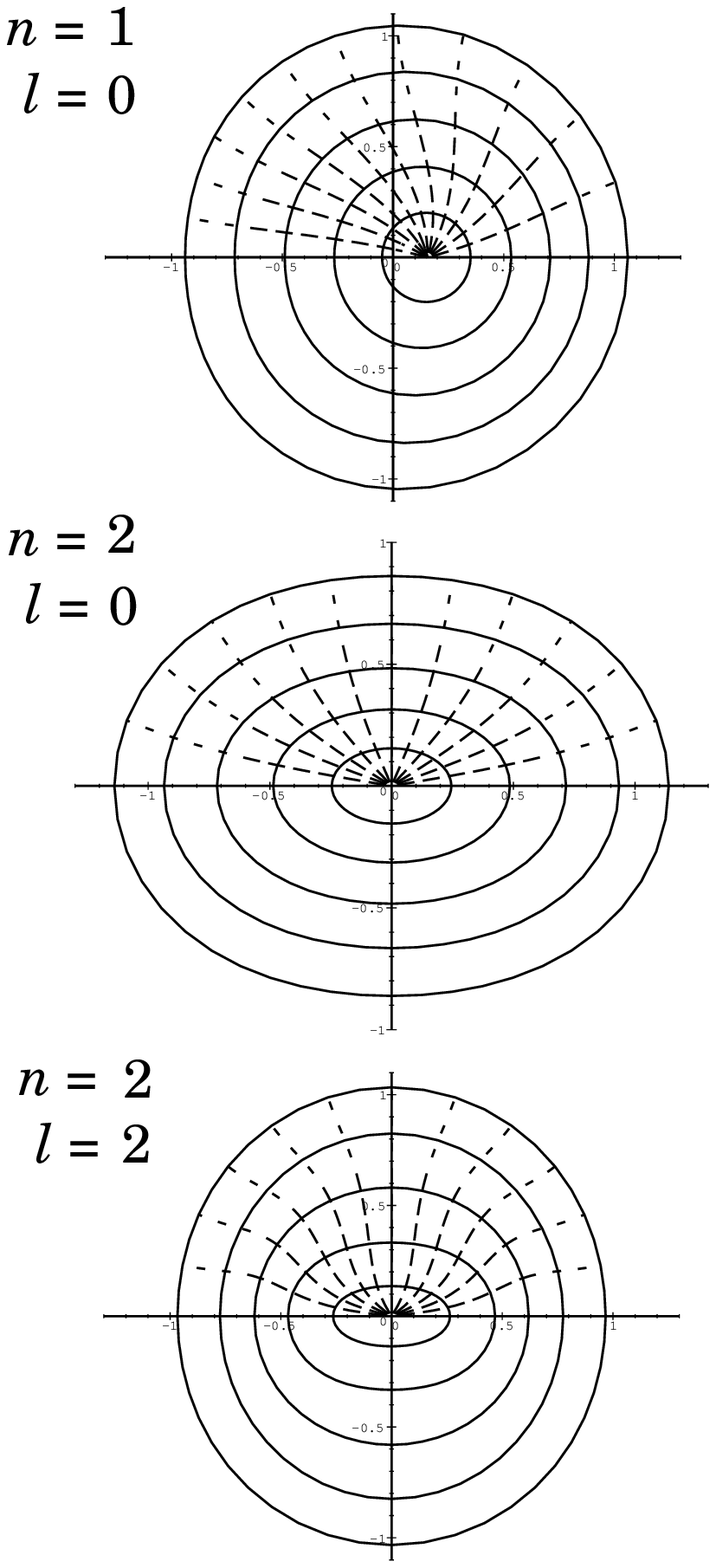}
\caption[Eigenmodes]
{}
\label{wavevgl}
\end{figure}

\end{document}